\documentstyle[psfig,namedreferences]{kluwer}
\def\arcdeg{\ifmmode^\circ\else$^\circ$\fi}

\runningtitle{ EVOLUTION  OF  THE  MAGNETIC   FIELD IN  AR5747}
\runningauthor{YONG-JAE MOON ET AL.}
\begin{opening}
\title{
EVOLUTION OF THE MAGNETIC FIELD IN  
AR 5747 AND ITS APPROXIMATION 
AS A LINEAR FORCE-FREE FIELD}
\author{Y.-J. \surname{MOON}\thanks{
e-mail:yjmoon@boao.re.kr}}
\institute{Korea Astronomy Observatory, Whaamdong, Yooseong-ku, Taejon, 
305-348, Korea}
\author{H. S. \surname{YUN}}
\institute{Department of Astronomy, Seoul National University, Seoul 151-742,
Korea}
\author{G. S. \surname{CHOE}}
\institute{ Princeton Plasma Physics Laboratory, Princeton, NJ 08543-0451, USA}
\author{Y. D. \surname{PARK}}
\institute{Korea Astronomy Observatory, Whaamdong, Yooseong-ku, Taejon, 305-348, Korea}
\author{D. L. \surname{MICKEY}}
\institute{Institute for Astronomy, University of Hawaii, 2680
Woodlawn Drive, Honolulu, HI 96822-1839, USA}
\date{}
\end{opening}
\begin{document}
\begin{abstract}
The evolution of nonpotential characteristics of magnetic
fields in AR 5747 is presented using 
Mees Solar Observatory magnetograms taken on 
Oct. 20, 1989 to Oct. 22, 1989. 
The active region showed such violent flaring activities
during the observational span  
that strong X-ray flares took place including a 2B/X3 flare.
The magnetogram data were obtained 
by the Haleakala Stokes Polarimeter which provides
simultaneous Stokes profiles of the
Fe I doublet 6301.5 and 6302.5. A nonlinear least square
method was adopted to derive the magnetic field vectors from the
observed Stokes profiles and a multi-step ambiguity solution
method was employed to resolve the
$180\arcdeg$ ambiguity. 
From the ambiguity-resolved vector magnetograms,
we have derived a set of physical quantities characterizing 
the field configuration, which are magnetic flux, vertical current
density, magnetic shear angle, angular shear, magnetic free energy
density and a measure of magnetic field discontinuity MAD
(Maximum Angular Difference between two adjacent field vectors).
In our results, all the physical parameters decreased with time, 
which implies that the active region 
was in a relaxation stage of its evolution. 

To examine the force-free characteristics of the field, 
we calculated the integrated Lorentz force and 
and also compared the longitudinal field 
component $B_z$ with the corresponding 
vertical current density $J_z$.  In this investigation, we found 
that the magnetic field in this active region was approximately 
linearly force-free 
throughout the observing period.
The time variation of the linear force-free
coefficient is consistent with
the evolutionary trend of other nonpotentiality  parameters.
This suggests that  
the linear force-free coefficient 
could be a good indicator of the evolutionary 
status of active regions. 
\end{abstract}

\section{Introduction}
It is generally believed that  magnetic fields play a central role in
solar eruptive phenomena such as flares and coronal mass ejections. 
The energy released through solar eruptive processes 
is considered to be stored 
in nonpotential magnetic fields. The magnetic energy is supplied 
to the corona either by plasma flows moving around magnetic
fields in the inertia-dominant photosphere or by magnetic flux 
emerging from below the photosphere. 
Since measurements of magnetic fields at the coronal altitude 
are not  available, magnetograms taken at the photospheric level 
have been widely used
for studies of magnetic nonpotentiality in
flare-producing active regions and are also used through 
extrapolation to 
compute coronal magnetic fields.  

Several attempts have been made to identify the relationships between
time variation of 
nonpotentiality parameters and development of  solar flares 
(Hagyard {\it et al.}, 1984; Hagyard 
{\it et al.}, 1990; Wang {\it et al.}, 1996; Wang 1997).
Moon {\it et al.} (1999c, Paper I) reviewed previous studies on magnetic 
nonpotentiality 
indicators and discussed the problems involved in them. 
Specifically, they studied the evolution of nonpotentiality 
parameters in the course of 
an X-class flare of AR 6919 using  MSO (Mees Solar Observatory) magnetograms. 
They showed that the magnetic shear obtained from
the vector magnetograms increased just before the flare and then decreased after it, 
at least near the $\delta$ spot region. 
Moon {\it et al.} (1999a) proposed 
a measure of magnetic field discontinuity, MAD, 
defined as Maximum Angular Difference between two adjacent field vectors, 
as a flare activity indicator. They applied this concept 
to three magnetograms of AR 6919 and found that the high MAD regions well match
the soft X-ray bright points observed by Yohkoh. 
It was also found that the MAD values  increased  just before 
an X-class flare and then
decreased after it.
This paper constitutes one of the series of studies 
on evolution of magnetic nonpotentiality  associated with major X-ray flares,  which are 
performed using MSO vector
magnetograms.

Metcalf {\it et al.} (1995) studied MSO magnetograms of AR 7216, which are obtained 
from the observation of Na I 5896
spectral line,  employing 
a weak field derivative method (Jefferies and Mickey, 1991)
and concluded that the magnetic field of AR 7216 is 
far from force-free at the photospheric level. 
This method could underestimate  transverse field strengths for strong field regions 
due to the saturation effect and calibration problems
(Hagyard and Kineke, 1995; Moon, Park, and Yun, 1999b).  
Metcalf {\it et al.} (1991) used  polarization data of Fe I
 6302 line to compare two  calibration 
methods: the weak field derivative method (Jefferies and Mickey, 1991) and the nonlinear 
least square method (Skumanich and  Lites, 1987). They found that  there are noticeable 
differences in magnetic field strength larger than  $B \ge 1200 \, \rm G$ 
between the two methods. 

On the other hand, Pevtsov {\it et al.} (1997) analyzed 655 photospheric magnetograms of 140 
active regions to examine the spatial variation of  force-free coefficients. 
In their results, some of the active regions show a good correlation between $B_z$  and 
$J_z$, but others  do not.    It is  very natural  that the force-free  coefficient varies 
depending on 
the active region in question even when 
the coefficient is more or less constant over the 
active region. Now we raise the question of whether a certain relation
can be drawn 
between the force-free coefficient  
and the evolutionary stage of an active region. 
Among the selections by Pevtsov {\it et al.} (1997), AR 5747 is exemplary of 
showing a good correlation between $B_z$ and $J_z$. Thus, we have taken AR 5747 
as the object of our study. 

The purpose of this paper is to examine the magnetic nonpotentiality of AR 5747 
associated with solar flares and investigate the evolution of 
the linear force-free coefficient.  
For this study, we have used the magnetograms spanning three days  
obtained from full Stokes polarization profiles of MSO. 
In Section 2, an description is given of the 
observation and analysis of the vector magnetograms. 
Computation of   nonpotentiality parameters  and  
their evolutions   in relation  to  flaring 
activities is presented in Section 3. 
In Section 4,  we discuss the evolution of the active region field 
as a linear force-free field. 
Finally, a summary and conclusion are given in Section 5.

\section{Observation and Analysis}

For the present work, we have selected a set of MSO magnetograms of 
AR 5747 taken on Oct. 20--22, 1989.
The magnetogram data were obtained by the 
Haleakala Stokes polarimeter (Mickey, 1985) 
which provides simultaneous Stokes I,Q,U,V profiles  
of the Fe I 6301.5, 6302.5 \AA \  doublet. 
The observations were made by a rectangular raster scan
with a pixel spacing of  5.6$''$ (low resolution scan)
and a dispersion
of 25 ${\rm m \AA / pixel}$. Most of the analyzing procedure is  
well described in Canfield  {\it et al.} (1993).
To derive the magnetic field vectors from Stokes
profiles, we have used a nonlinear least square fitting 
method (Skumanich and Lites, 1987) for fields stronger than 
100 G and an integral method (Ronan, Mickey and
Orral, 1987)
for weaker fields.
In that fitting, 
the Faraday rotation effect, which is one of the error sources
for strong fields, is properly taken into account.
The noise level in the original magnetogram is about 70 G for
transverse fields and 10 G for longitudinal fields.
The basic observational parameters of the magnetograms used in this study are
presented in Table I.

To resolve the $180\arcdeg$ ambiguity,  
we have adopted a multi-step ambiguity solution
method by Canfield {\it et al.} (1993) (for details, see the
Appendix of their paper). In the 3rd and 4th steps of their method,   
they have chosen the orientation of the transverse field 
that minimizes the angle
between neighboring field vectors and the field divergence 
$|\nabla \cdot {\bf B}|$.

\begin{table}
\caption{Basic observational parameters of AR 5747. }
\begin{tabular}{cccccc}
\hline
Data  &  Date &  Time  & 
Scan &  Data Points & Coord. \\
\hline
AR5747 a) &  20 Oct., 1989 & 17:41-18:50 & 5.656$''$ & 30$\times$30 & S26W07\\
AR5747 b) &  21 Oct., 1989 & 19:20-20:16 & 5.656$''$ & 30$\times$30& S26W22 \\
AR5747 c) &  22 Oct., 1989 & 18:27-19:41 & 5.656$''$ & 30$\times$40& S26W33 \\
\hline
\end{tabular}
\end{table}

\section{Evoultion of Magnetic Nonpotentiality}

In the active region AR 5747,  a number of flares took place 
including a 2B/X3 flare. 
In Table II, we summarize
some basic features of major X-ray flares during the observing period.

\begin{table}
\caption{Basic information of the major X-ray flares observed in AR 5747 
(Solar Geophysical Data). }
\begin{tabular}{lccccccc}
\hline
ID & Date  &  Start(UT)  &  End   & 
Max. &  Coord. & Optical Class & X-ray Class  \\
\hline
F1 & 20/10/89 &  21:30 & 22:03 & 21:34 & S26W11 & 1N & M1.4 \\
F2 & 21/10/89 &  01:53 & 02:06 & 01:55 & S28W09 & 1N & M2.4 \\
F3 & 21/10/89 & 06:40 &06:49 & 06:43 &S27W16 & 1N  & M1.9\\
F4 & 21/10/89 & 23:54 &24:00 &23:56& S28W22&  &   M3.1\\
F5 & 22/10/89 & 11:15&12:30 &11:21& S27W26 & SN&                       M1.5\\
F6 & 22/10/89 & 15:54&16:47&15:58 & S28W28 &SN&                       M1.3\\
F7 & 22/10/89 & 17:08&21:08&17:57&S27W31 &2B &                      X2.9\\
\hline
\end{tabular}
\end{table}

\begin{figure}
\psfig{figure=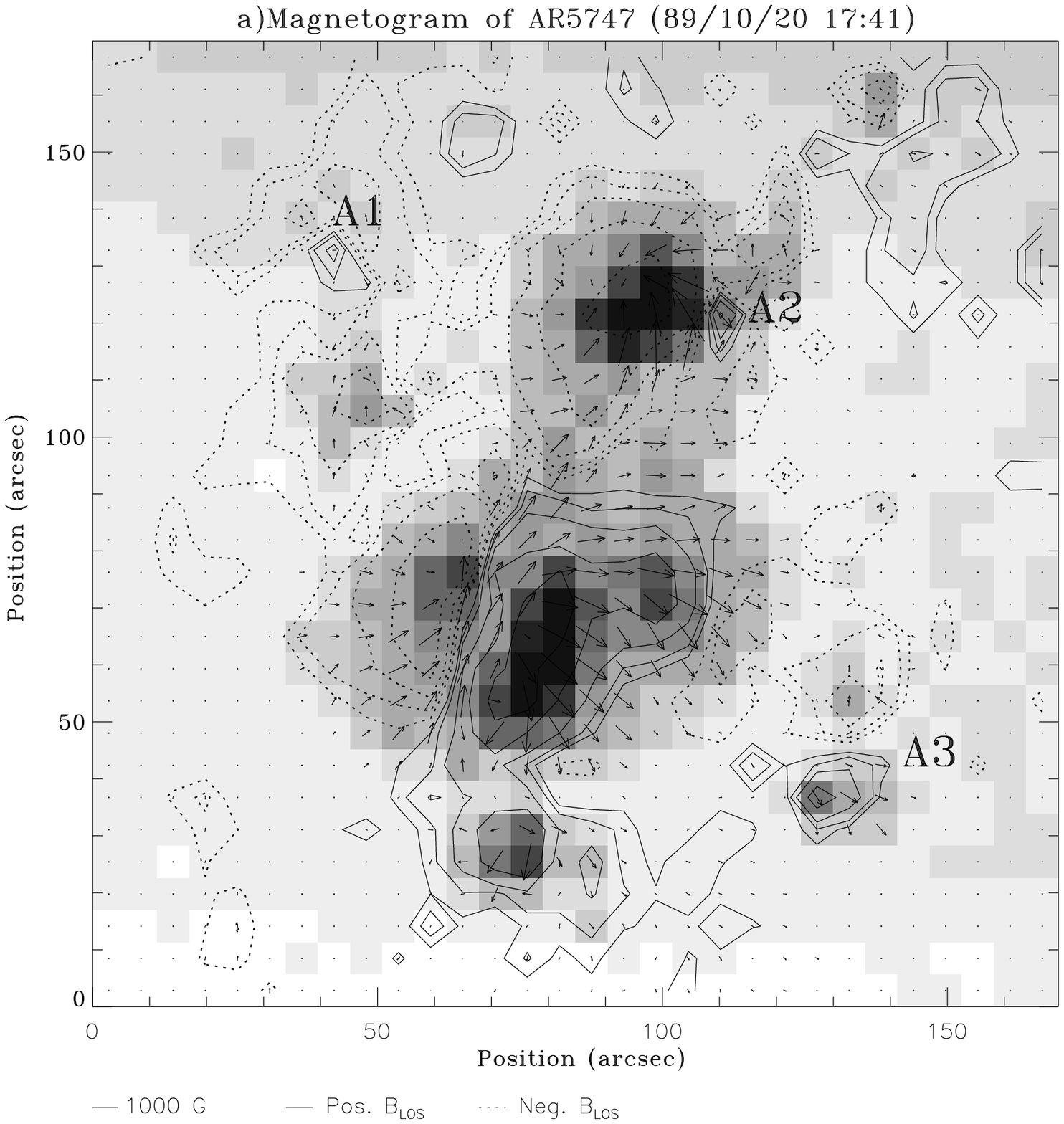,height=6cm,width=8cm}
\psfig{figure=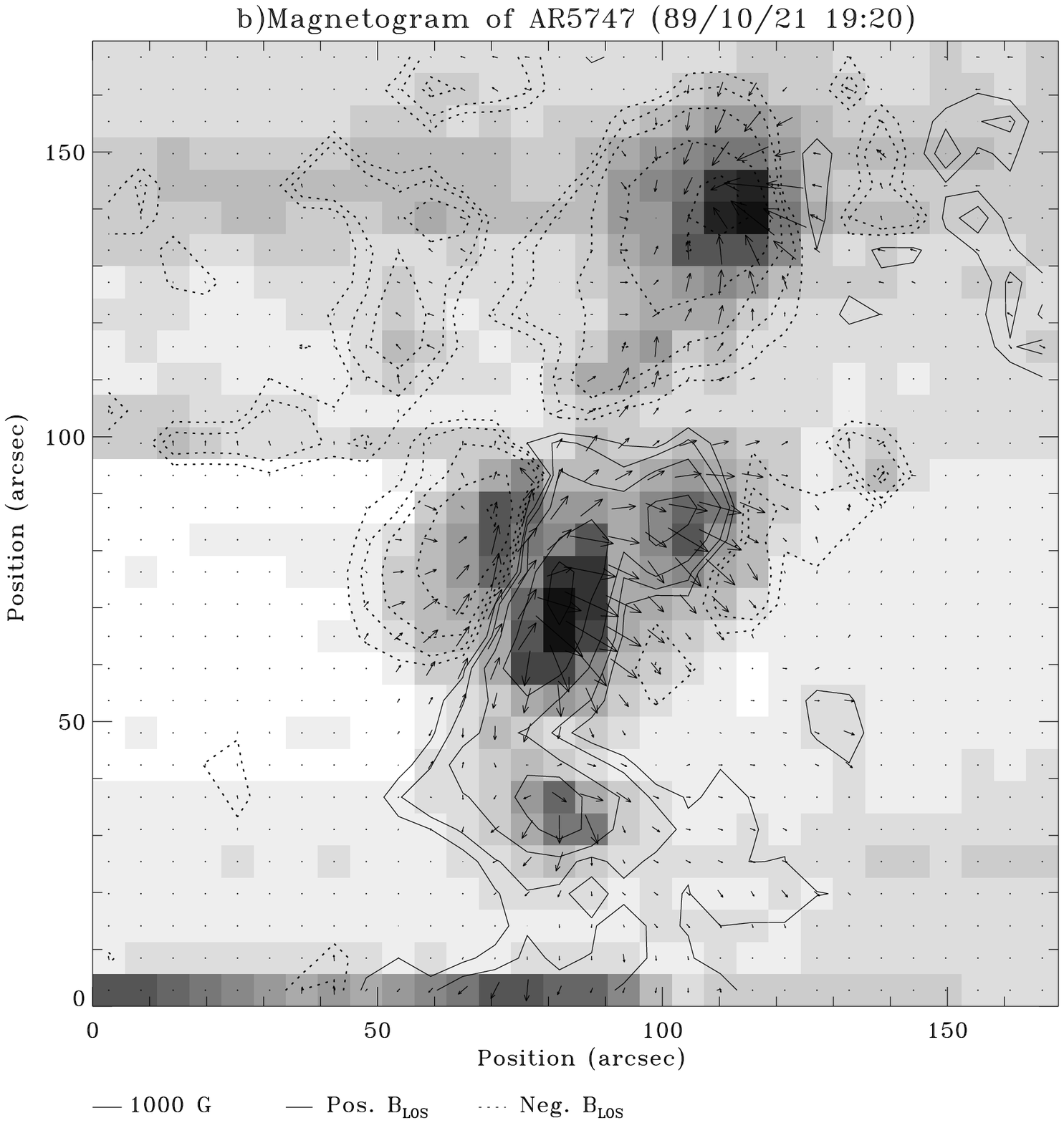,height=6cm,width=8cm}
\psfig{figure=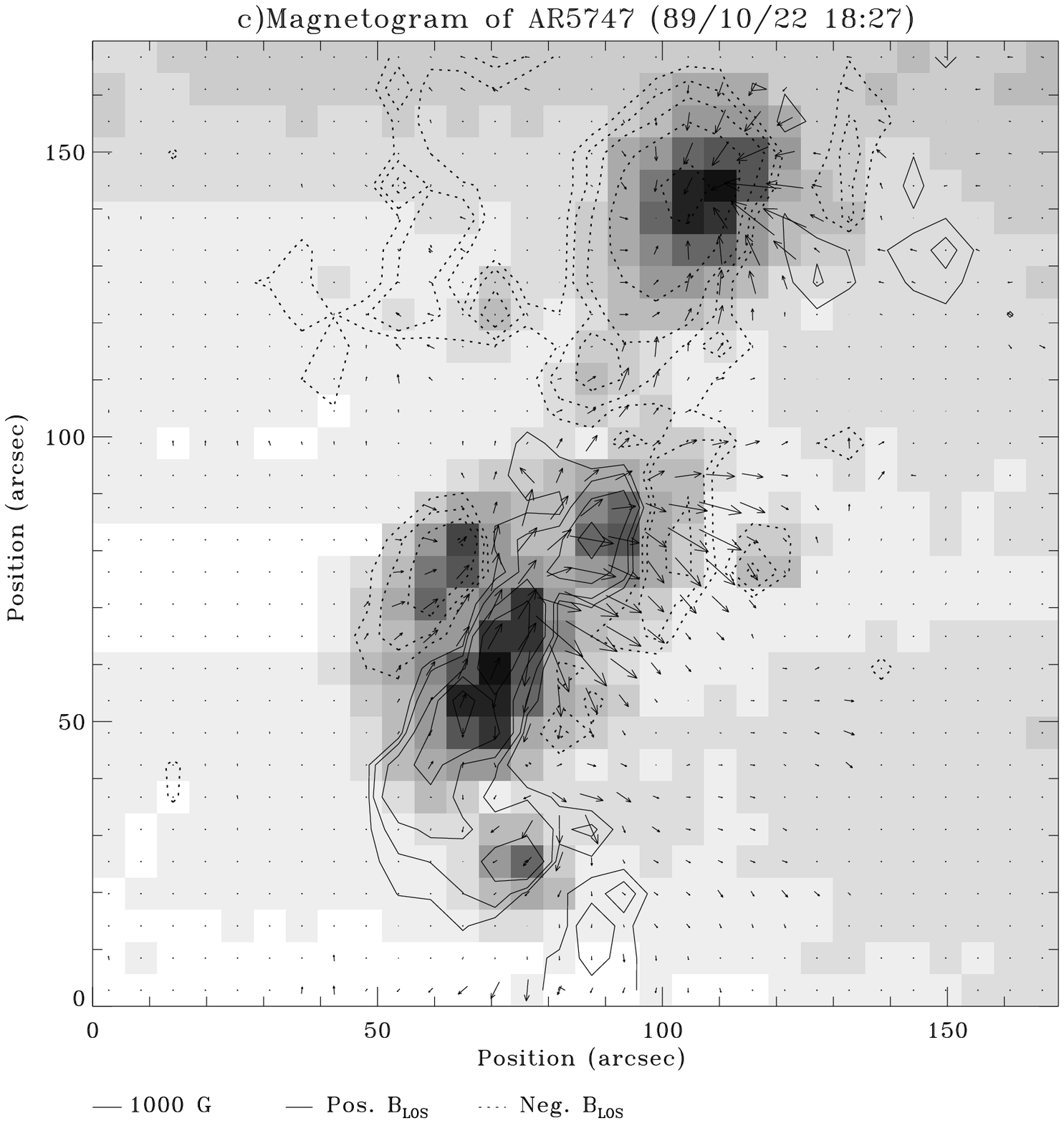,height=6cm,width=8cm}
\caption{
Three MSO vector magnetograms of AR 5747  superposed on
white light images taken on Oct. 20--22, 1991.
In all the figures, the  solid lines stand for    
positive polarities and  the   dotted lines for 
negative polarities. 
The contour levels correspond to 100, 200, 400, 800 
and 1600 G, respectively.    
The   length of arrows     represents   the magnitude of  the transverse 
field components. }
\end{figure}

Figure 1 shows the ambiguity resolved vector magnetograms obtained
on Oct. 20 to Oct. 22, 1989. The three magnetograms
have the same field of view. As seen in the figures, 
strong sheared transverse fields are concentrated near  the neutral line 
and they form a global clockwise winding pattern.  
In Paper I, an account is given of the magnetic nonpotentiality
parameters used in this study.
The vertical  current density is presented in Figure 2.  
The vertical current density 
kernels persisted, with little change of configuration, 
over the whole observing span. 
Wang, Xu, and Zhang (1994) and Leka {\it et al.} (1993) have 
discussed the important characteristics  
of these vector magnetic fields and vertical current 
densities.
We tabulate the time variation of magnetic fluxes
and  total vertical currents of positive and
negative signs in Table III. The  differences between the absolute values  of the positive 
and negative quantities are 
within a few percent.  
As seen in the  table, the magnetic fluxes  and total vertical 
currents of both signs decreased with time.
It is observed that several small $\delta$ sunspots
(A1, A2 and A3 in Fig. 1a) disappeared in Figure 1b, 
which suggests that flaring events 
between Oct. 20 and  21 should be associated with flux cancellation.
It is to be noted that there  were no  remarkable flux emergence  
during the  observing period.

\begin{table}
\caption{Magnetic fluxes and total vertical currents
in AR 5747 for three different times in the observing period. 
Here $\sigma_J$ denotes the standard deviation of the vertical current distribution.}
\begin{tabular}{cccccc}\hline
Data  & Flux(+) [Mx] &   
Flux(-) [Mx] &  $\sum J_z^{+}[\rm A]$ 
& $\sum J_z^{-} [\rm A]$  & $\sigma_J [\rm mA/m^2]$ \\
\hline
a) &  1.56E22 & 1.51E22 & 6.7E4 & 6.6E4 & 1.4\\
b) &  1.31E22 & 1.28E22 & 4.2E4 & 4.2E4 & 1.2\\
c) &  0.92E22 & 0.89E22 & 3.9E4 & 4.0E4 & 1.2\\
\hline
\end{tabular}

\end{table}

\begin{figure}
\psfig{figure=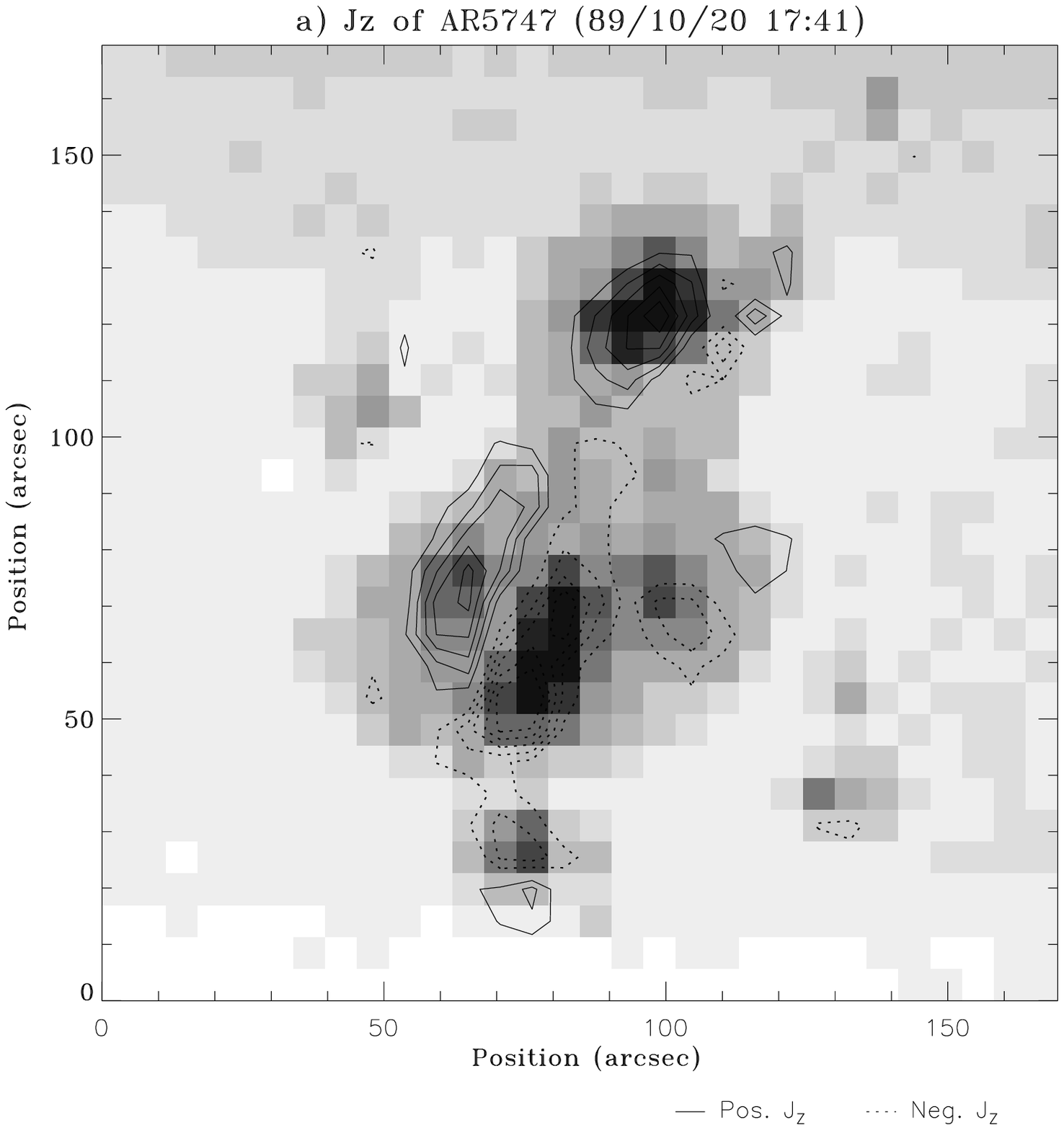,height=6cm,width=8cm}
\psfig{figure=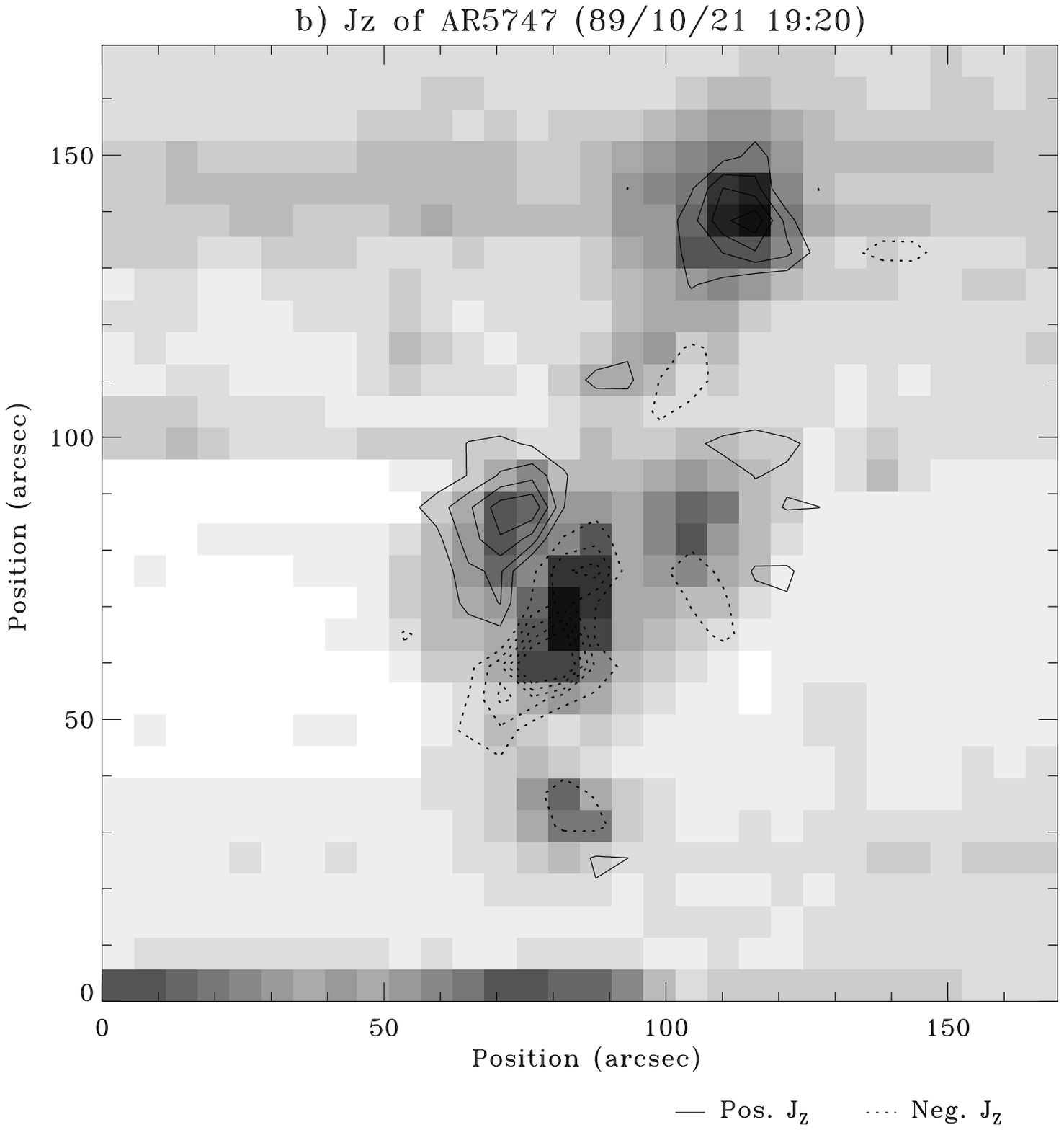,height=6cm,width=8cm}
\psfig{figure=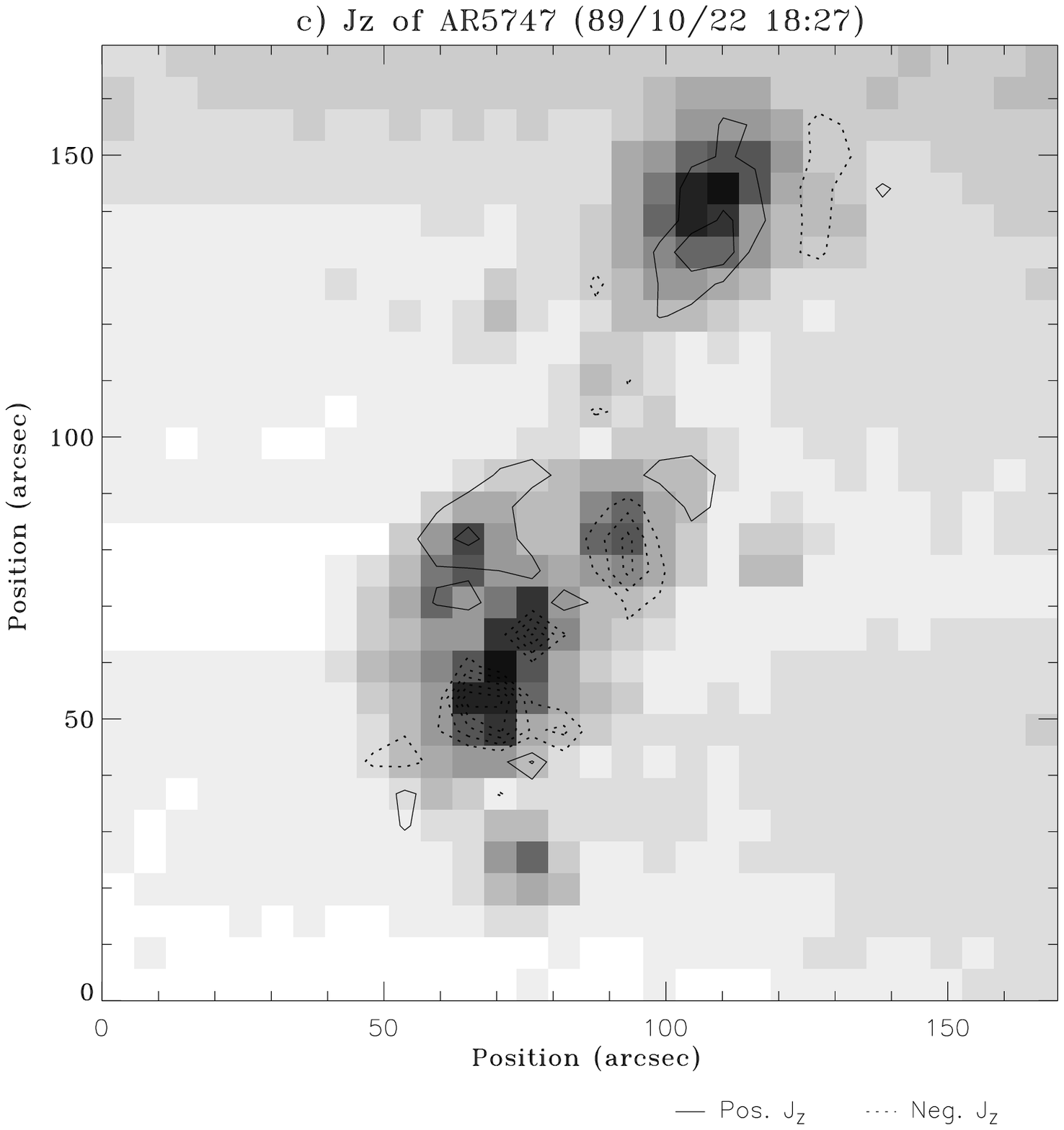,height=6cm,width=8cm}
\caption{ Contours of vertical current density   
in AR 5747. 
The contour levels correspond to
3, 6, 9, 12 and 15  $\rm mA/m^2$,
respectively.
In all the panels, the solid lines stand for positive
vertical current density and the dotted lines for 
negative current density.  
}
\end{figure}

\begin{table}
\caption{ Field strength weighted mean
magnetic shear angle $\overline \theta_s$, transverse field weighted angular 
shear $\overline \theta_a$, mean free energy density
$\overline \rho_f$, planar sum of magnetic free energy density $\sum \rho_f$,
and $\sum ({\rm MAD} \times |{\bf B}|)$
for AR 5747 for three different times. The values in parentheses are obtained 
employing the potential field method.}
\begin{tabular}{cccccc}
\hline
Data  & $\overline \theta_s$ &   
$\overline \theta_a$ &  $\overline \rho_f$ 
& $\sum \rho_f$  & $\sum ({\rm MAD} \times |{\bf B}|)$  \\
\hline
a) &  46.8(38.4) & 71.8(56.1) & 2.7(2.0)E4 & 1.4(1.0)E24 &4.7E6\\
b) &  41.3(36.3) & 63.0(53.4) & 2.1(1.6)E4 & 9.3(6.9)E23 & 3.5E6\\
c) &  36.1(32.6) & 55.7(48.3) & 1.7(1.4)E4 & 5.1(4.3)E23 & 2.3E6\\
\hline
\end{tabular}

\end{table}

\begin{figure}
\psfig{figure=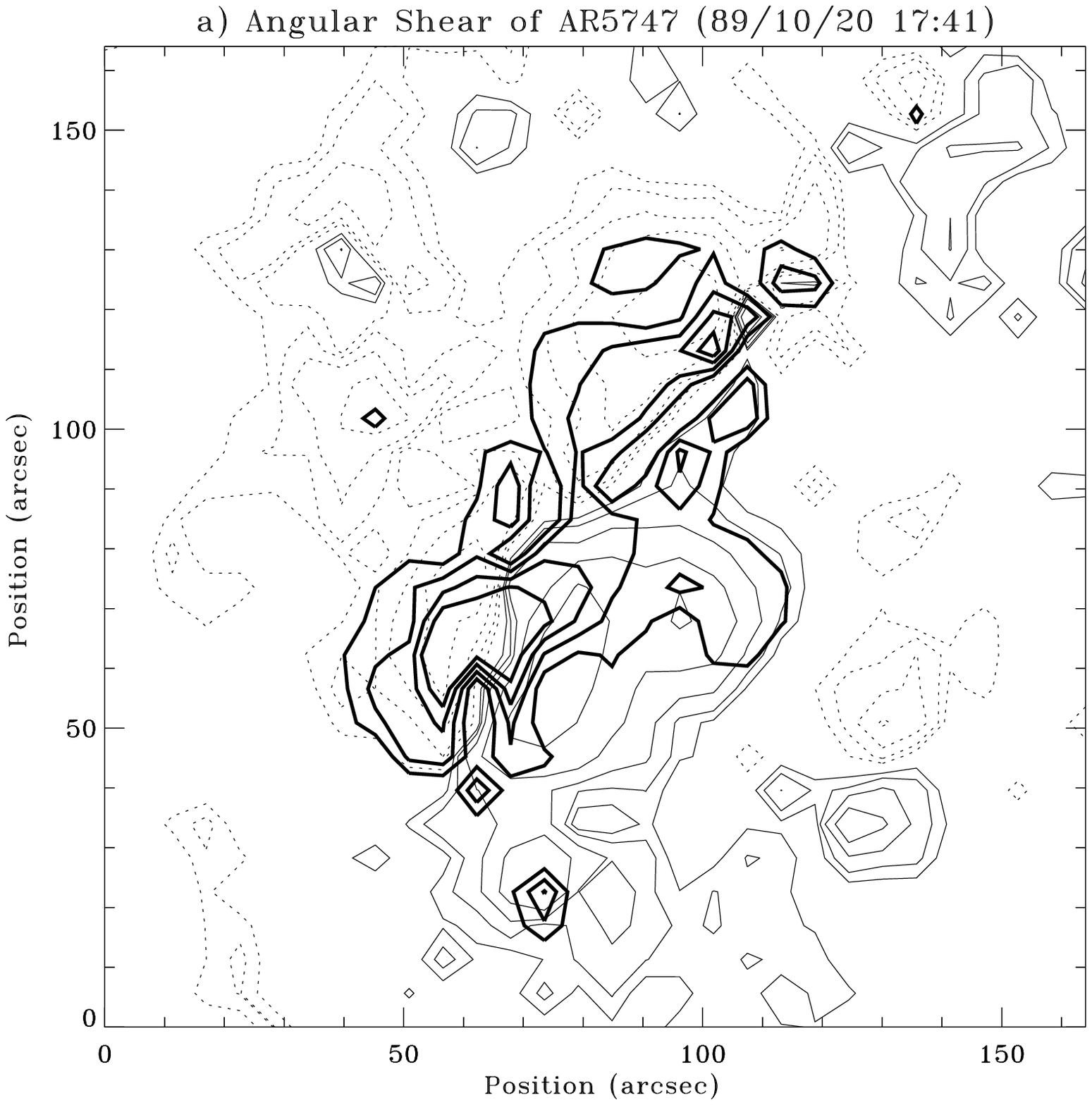,height=6cm,width=8cm}
\psfig{figure=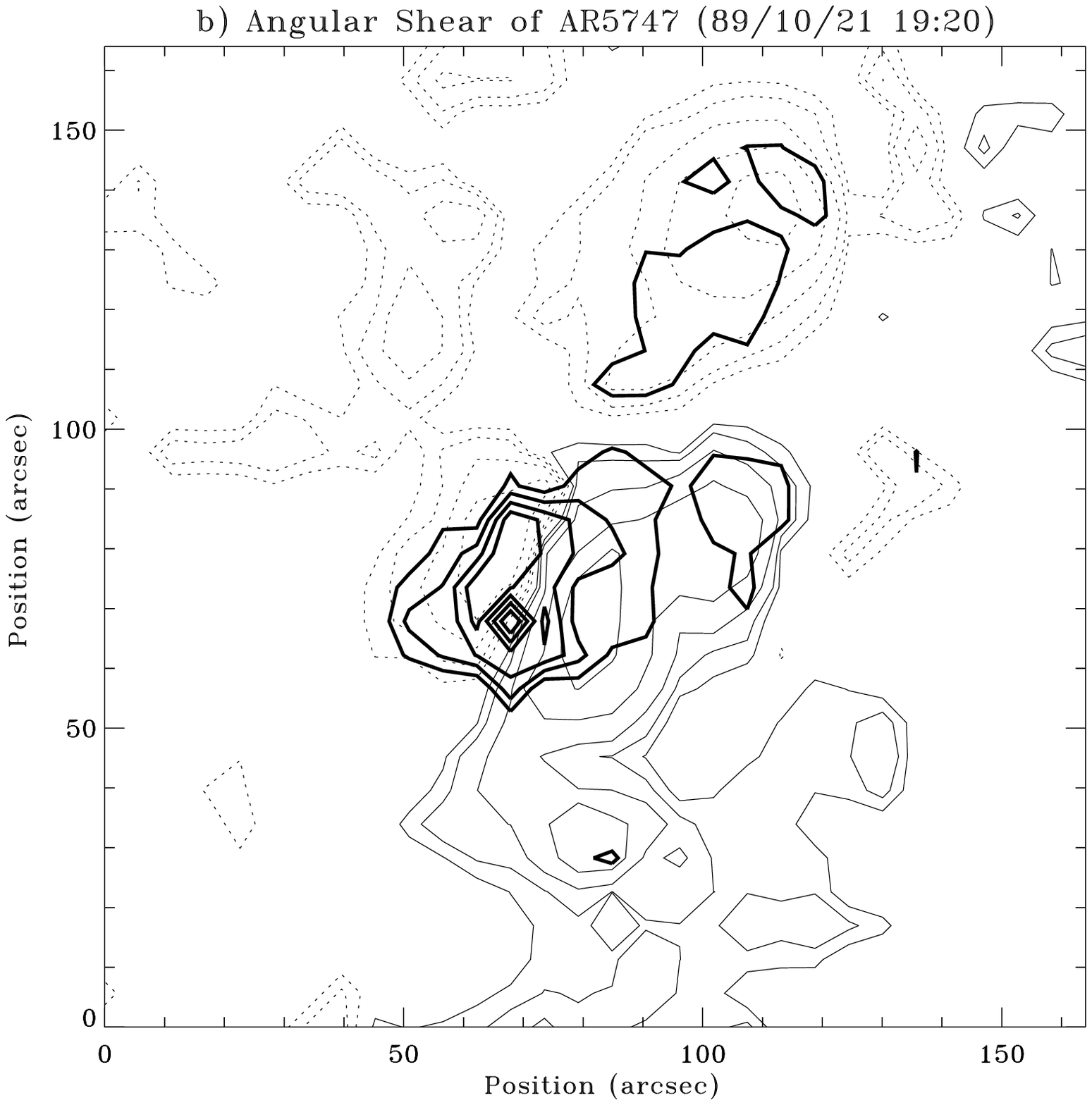,height=6cm,width=8cm}
\psfig{figure=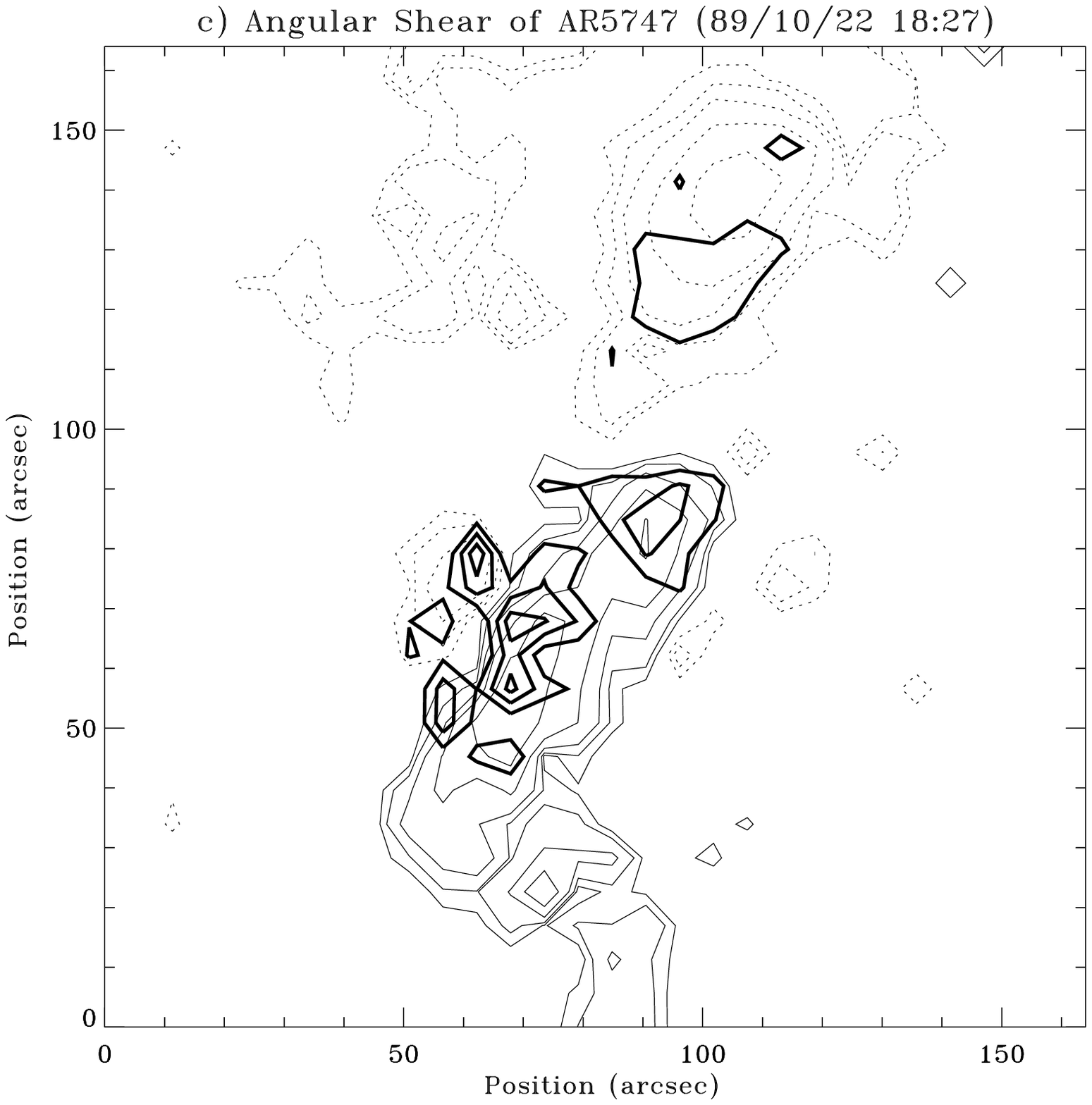,height=6cm,width=8cm}
\caption{ Contours of angular shear multiplied by transverse field strength 
drawn in thick solid lines are superposed on longitudinal magnetograms. 
The contour levels are $4.0 \times 10^4$, $7.0 \times 10^4$, $1.0 \times 10^5$
and $1.3 \times 10^5 \, \rm G\, deg$, 
respectively.
In all the panels, the solid lines stand for positive
longitudinal polarities and the dotted lines for 
negative ones. 
The contour levels in the magnetograms correspond to  
100, 200, 400, 800 and 1600 G,
respectively.
}
\end{figure}

\begin{figure}
\psfig{figure=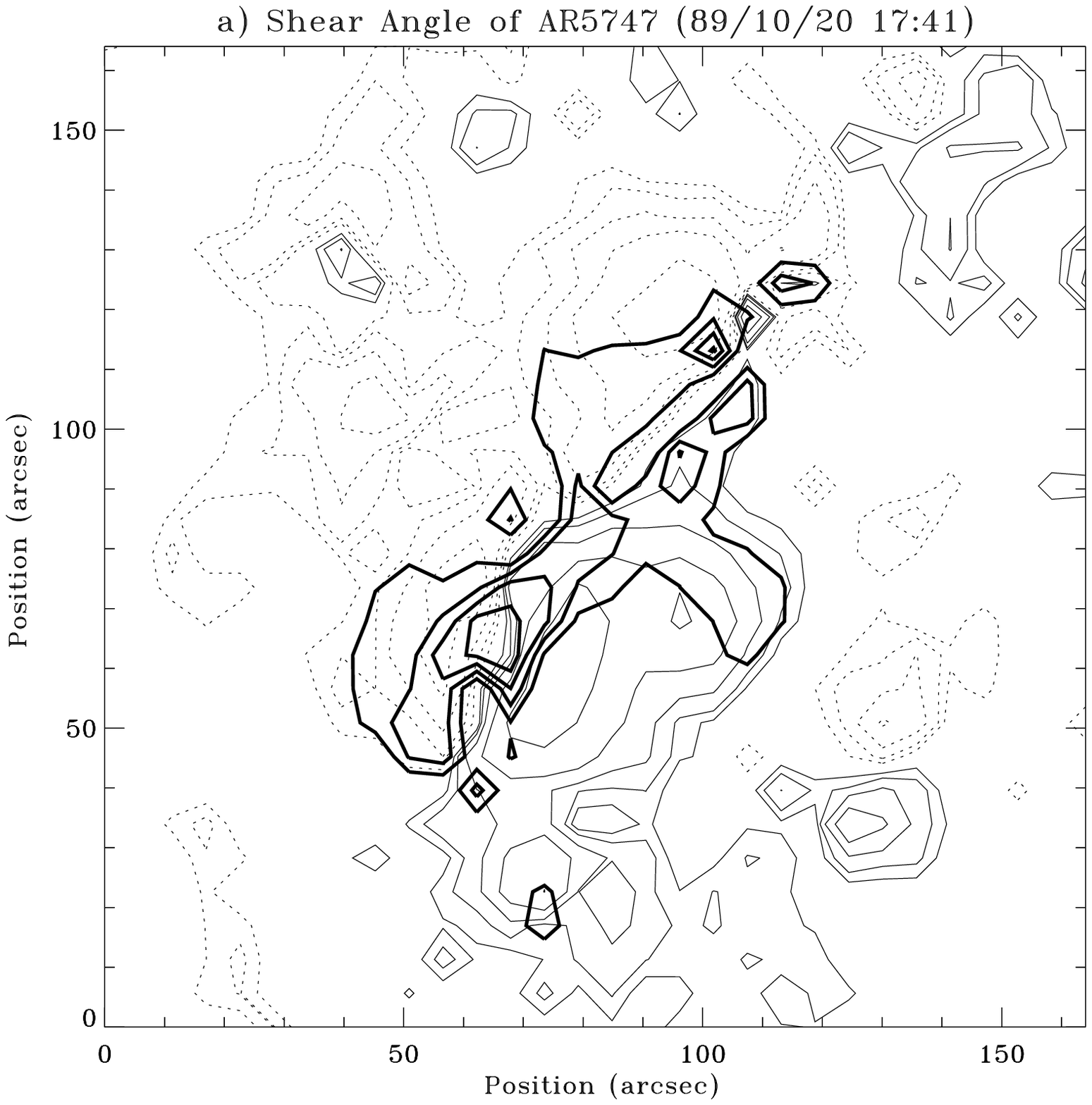,height=6cm,width=8cm}
\psfig{figure=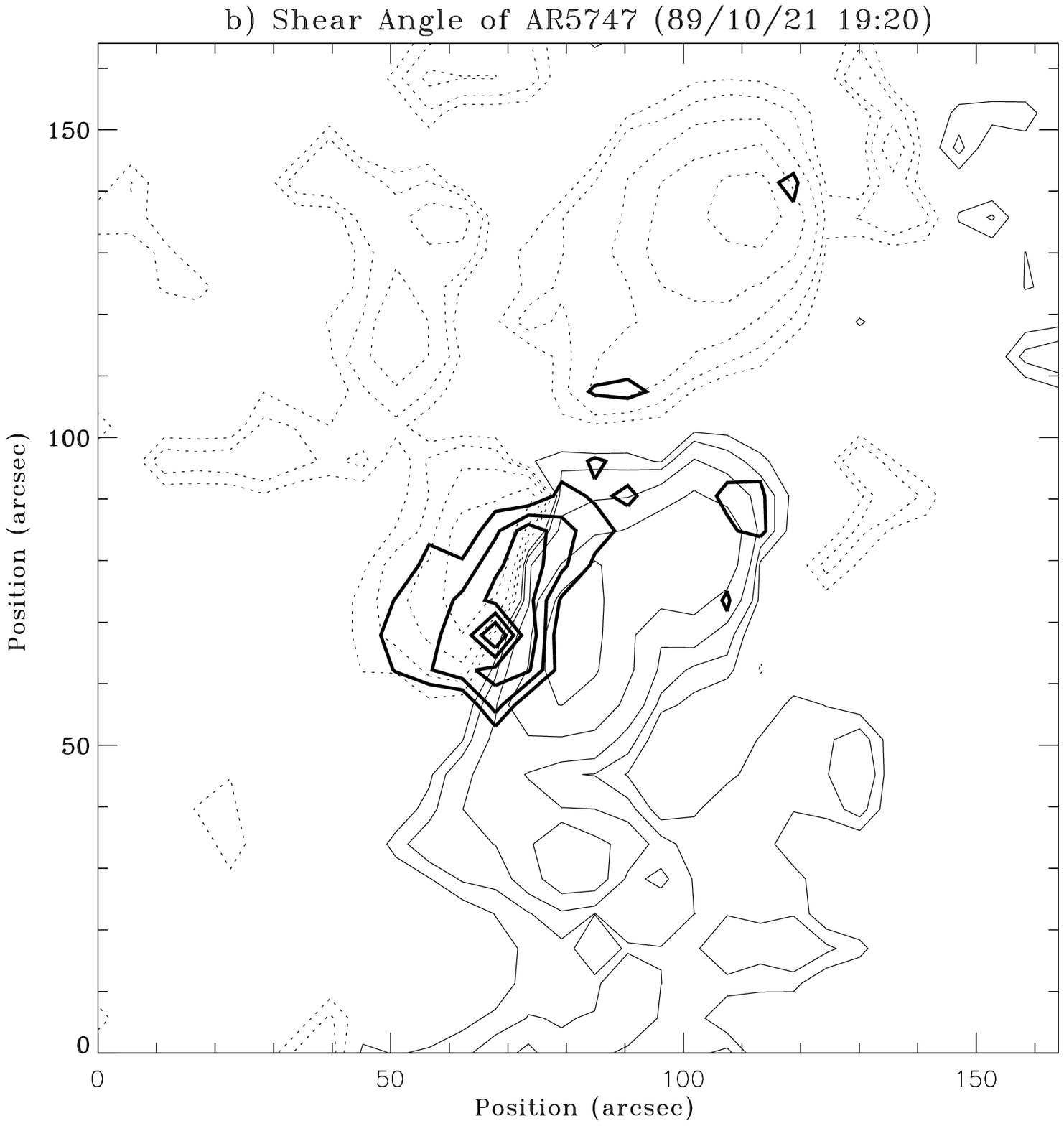,height=6cm,width=7.5cm}
\psfig{figure=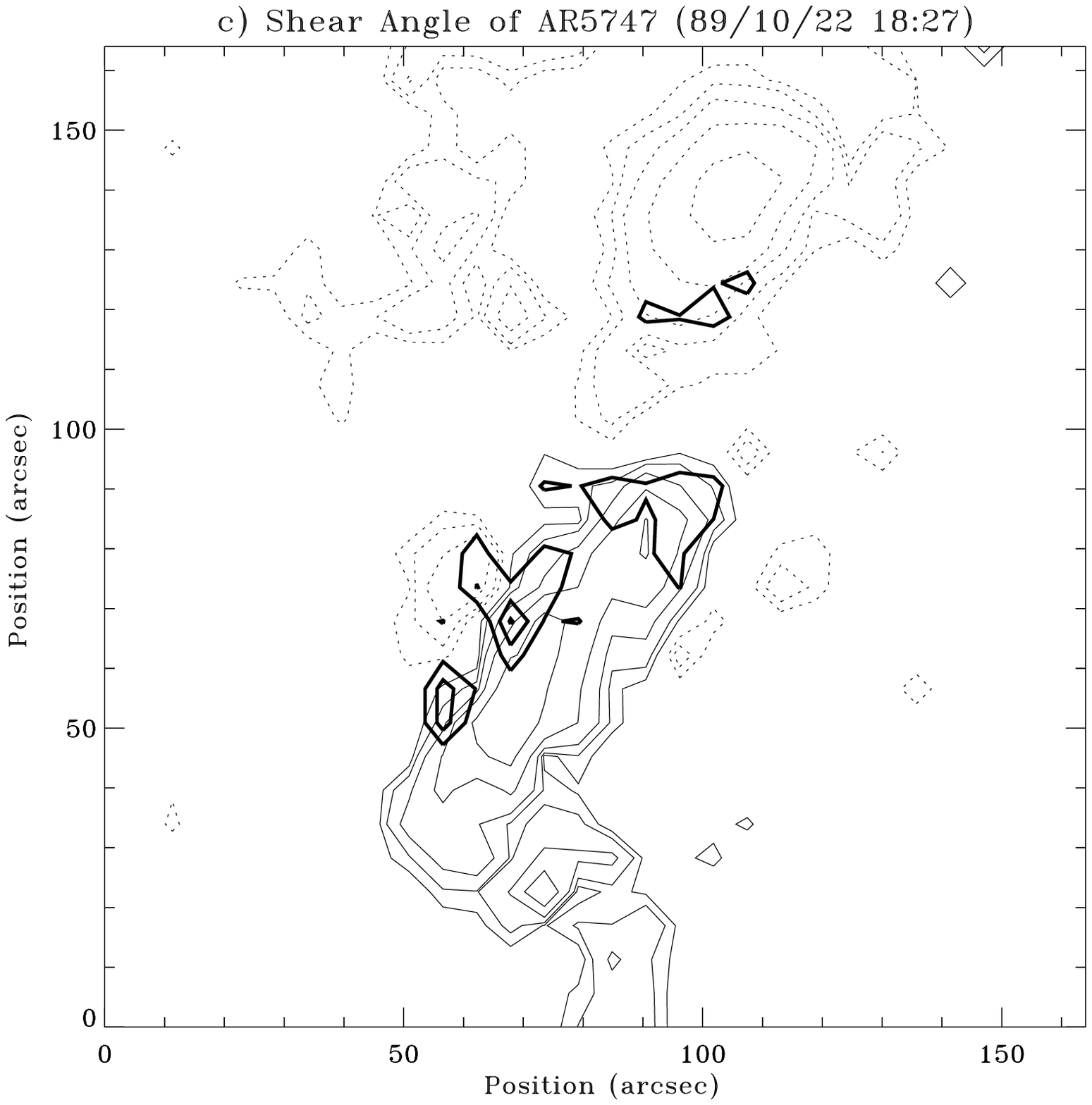,height=6cm,width=8cm}
\caption{Contours of shear angle
multiplied by field strength 
drawn in thick solid lines 
are superposed on longitudinal magnetograms.  
The contour levels are $4.0 \times 10^4$, $7.0 \times 10^4$, $1.0 \times 10^5$
and $1.3 \times 10^5 \, \rm G\, deg$, 
respectively.
The magnetograms are the same as in Figure 3.}
\end{figure}

\begin{figure}
\psfig{figure=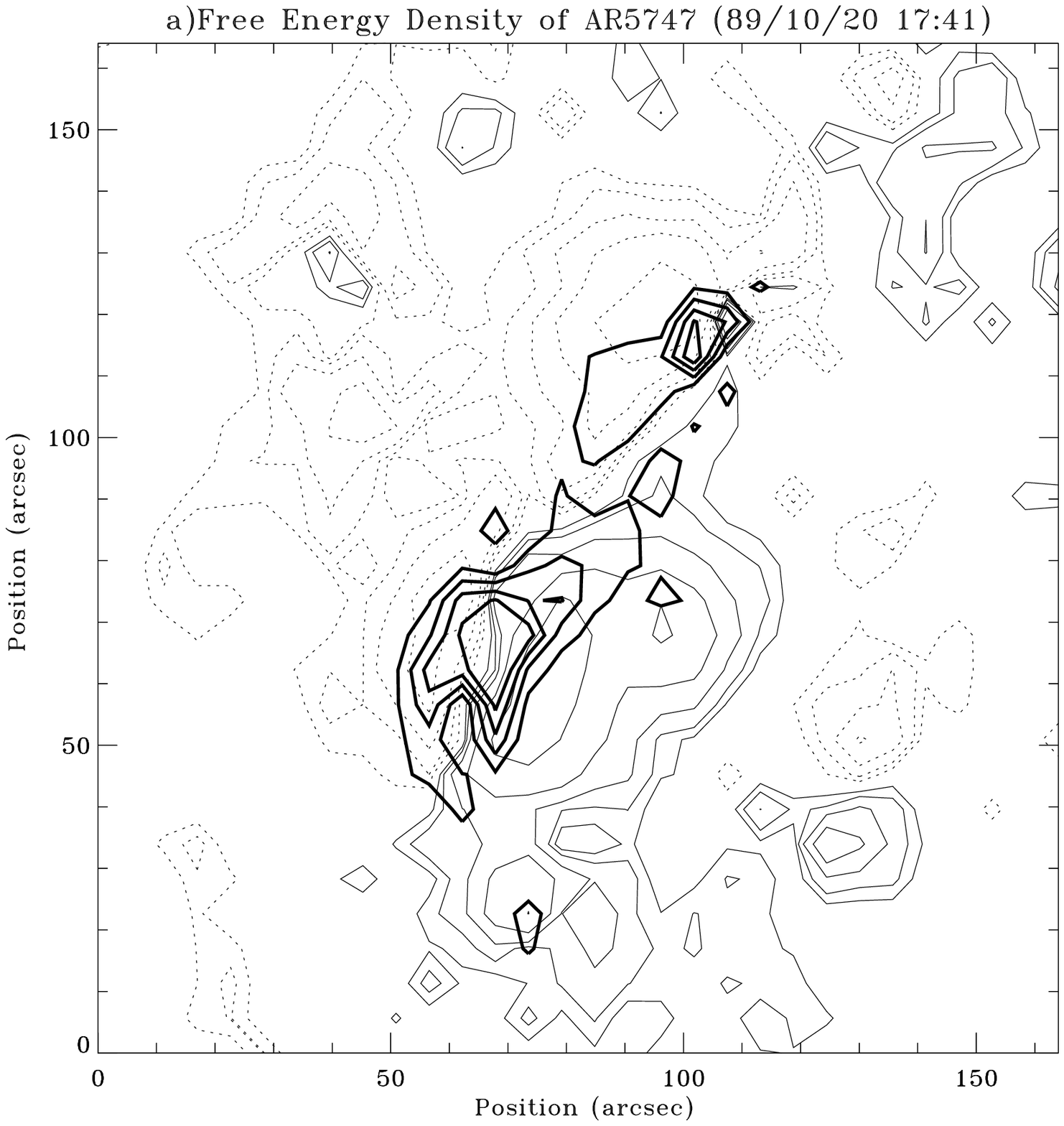,height=6cm,width=8cm}
\psfig{figure=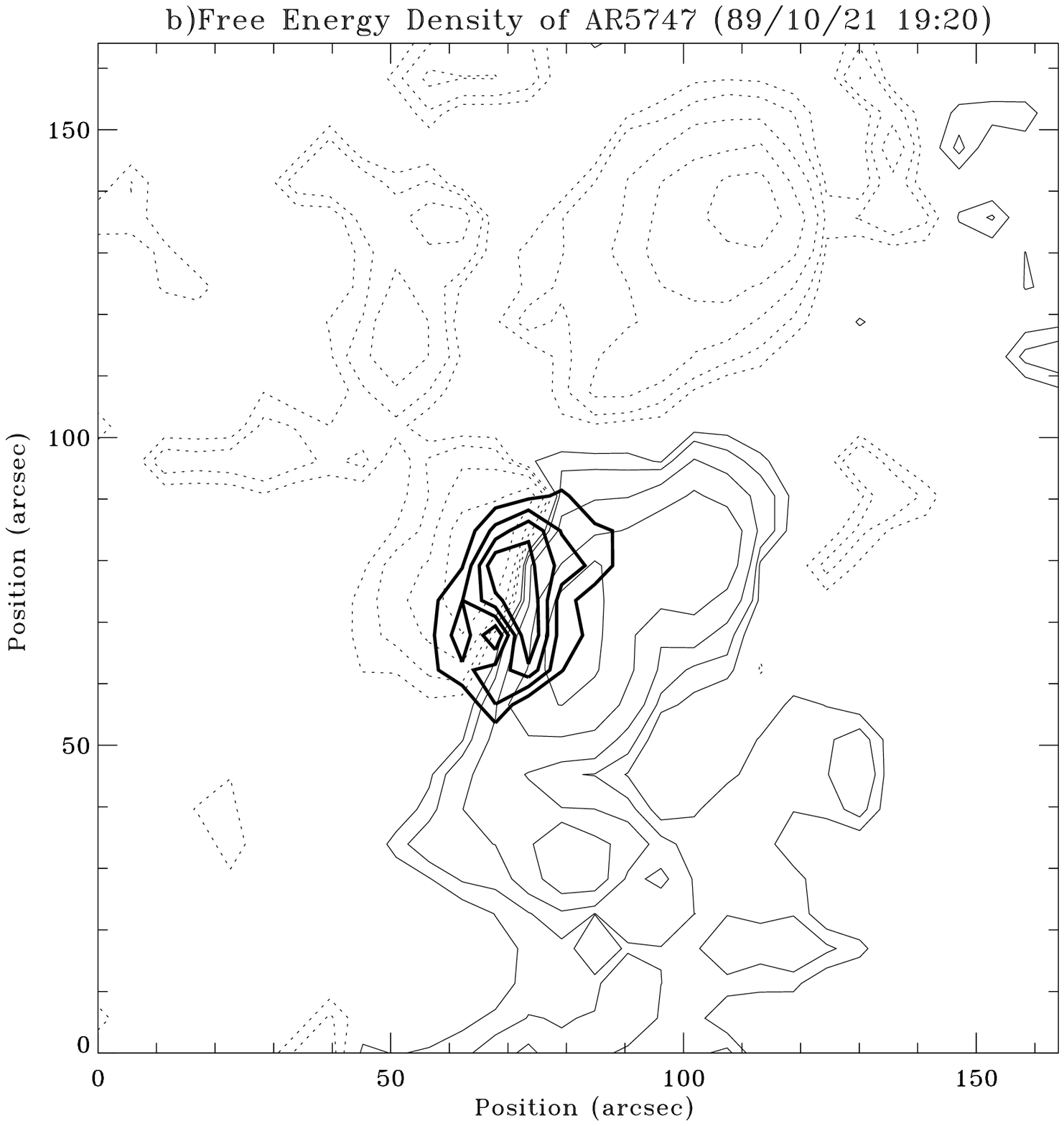,height=6cm,width=8cm}
\psfig{figure=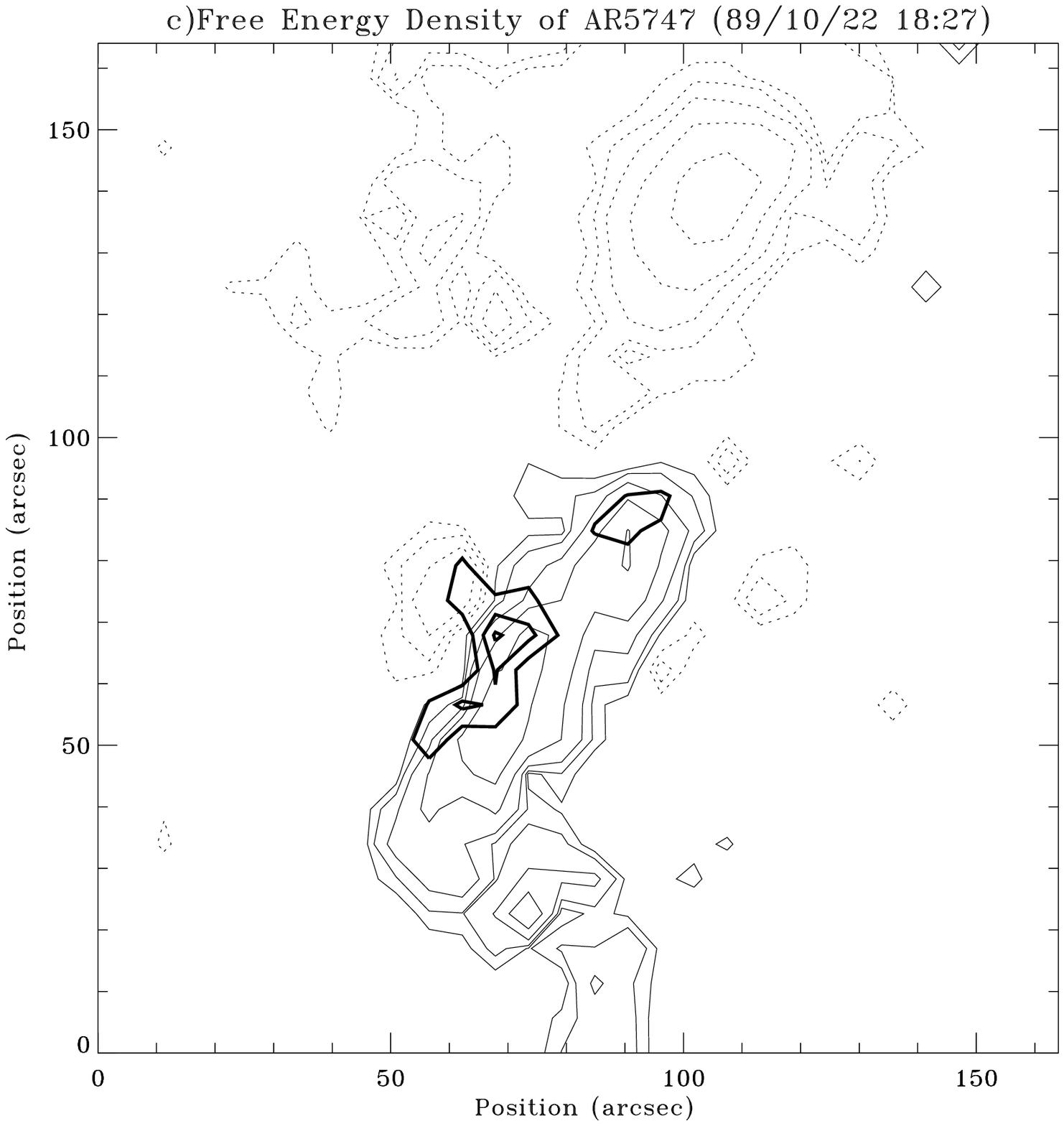,height=6cm,width=8cm}
\caption{Contours of free energy density drawn in thick solid lines are 
superposed on magnetograms. 
The contour levels are $5.0 \times 10^4$, $1.0 \times 10^5$, $1.5 \times 10^5$,
$2.0 \times 10^5$ and $2.5 \times 10^5 \, \rm erg/cm^3$, 
respectively.
The magnetograms are the same as in Figure 3.}
\end{figure}

\begin{figure}
\psfig{figure=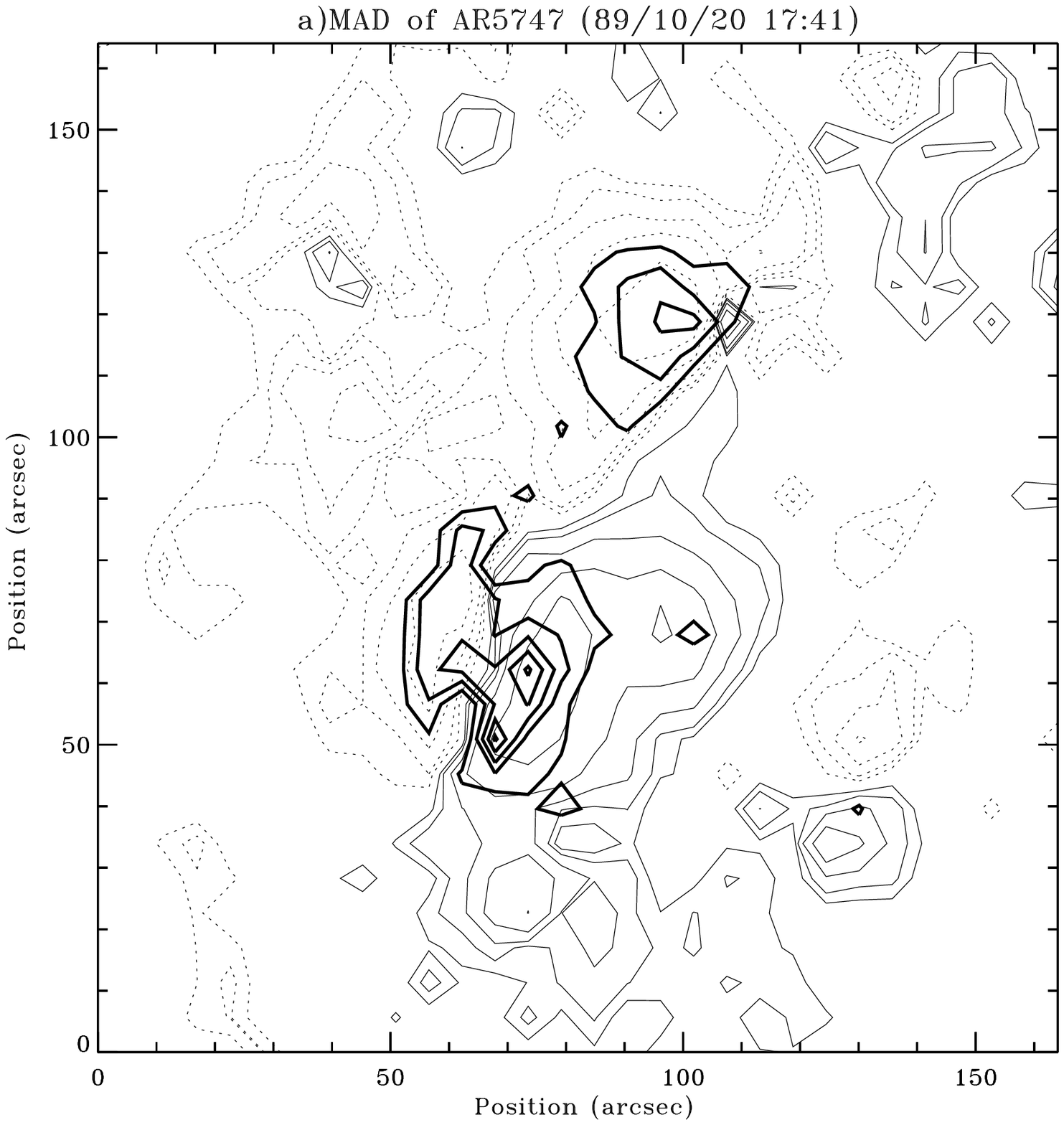,height=5.8cm,width=8cm}
\psfig{figure=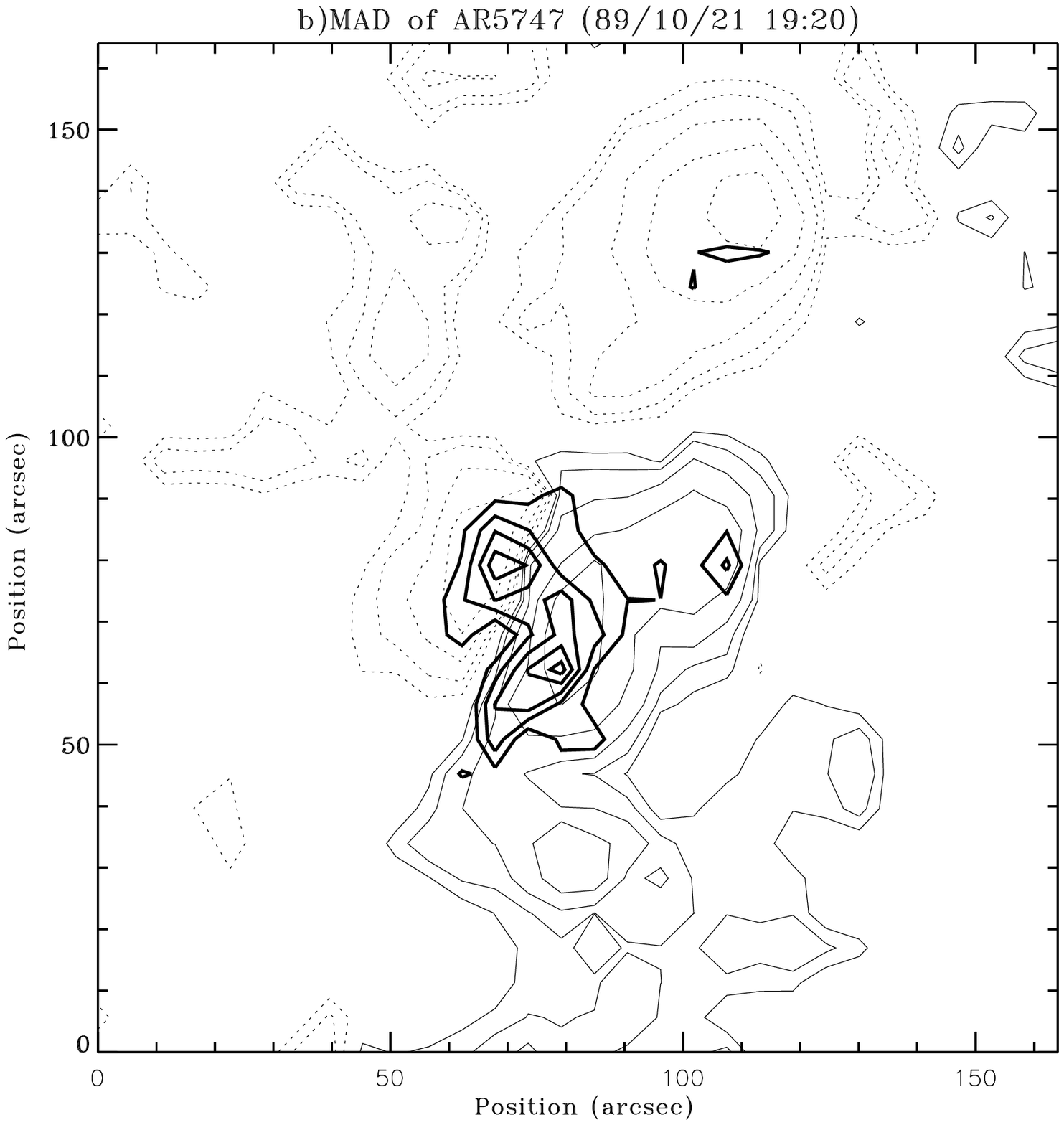,height=6cm,width=8cm}
\psfig{figure=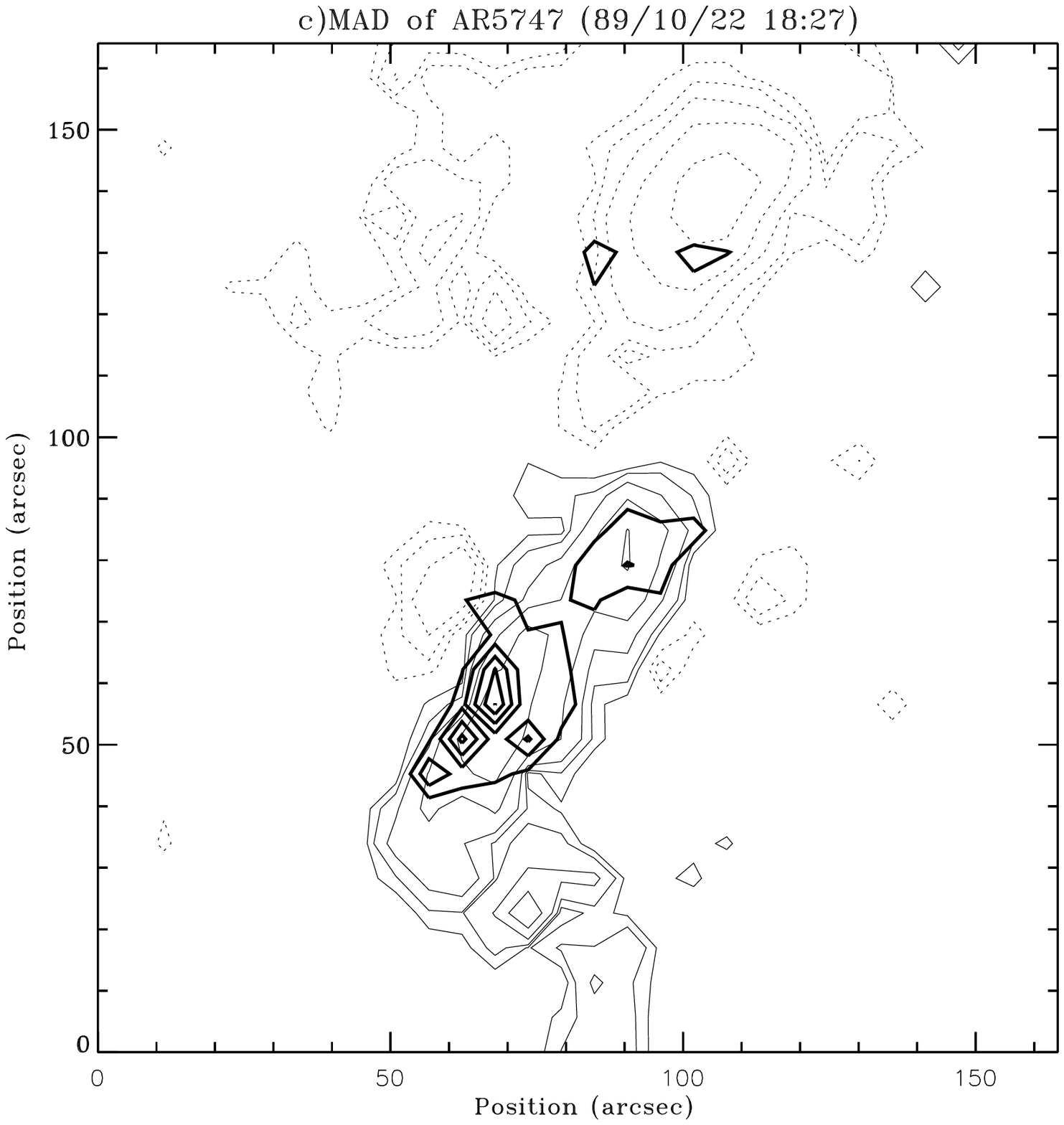,height=6cm,width=8cm}
\caption{ Contours of MAD multiplied by field strength 
drawn in thick solid lines 
are superposed on longitudinal magnetograms.  
The contour levels are $3.0 \times 10^4$, $5.0 \times 10^4$, $7.0 \times 10^4$,
$9.0 \times 10^4$ and $1.1 \times 10^5 \, \rm G\, deg$, 
respectively.
The magnetograms are the same as in Figure 3.}
\end{figure}

Figure 3 shows the  angular shear multiplied by transverse field strength and Figure 4
shows the shear angle multiplied by total field strength. 
As seen in the figures, strong magnetic shear is concentrated near
the inversion line,  where ${\rm H}_\beta$ emission patches were observed 
(see Fig. 2 of 
Wang, Xu, and Zhang 1994).
The time variation of two weighted
mean shear angles  is given in  Table IV. The values of two shear angles 
monotonically decreased  with time.  
The magnetic free energy density
is shown in  Figure 5.  Its evolutionary trend is quite similar to that of shear angles. 
The 2-D MAD  multiplied by  total field strength (Figure 6) also has a similar 
evolutionary pattern to that of the other nonpotentiality parameters above.  
We summarize the variations of mean free energy density, planar sum of free energy
density, and sum of MAD multiplied by field strength in Table IV, 
in which the values obtained with the potential field method for the $180\arcdeg$ 
ambiguity resolution are also given in parentheses for 
comparison.
As seen in the table, all the nonpotentiality parameters under consideration
decreased with time, which suggests that the active region was in a 
relaxation stage during the observing period. 

From the above results, we may infer that the flares that occurred in our
observation are just bursty parts of energy release in a long term relaxation of 
the stressed magnetic field. In a self-organizing system, a transition 
toward a lower energy state proceeds very mildly in the beginning and for 
most of time until a sudden 
bursty event develops as in an avalanche.  Why, then, did a series of flares occur, rather
than one? Flares can surely take place in repetition if enough energy is supplied into
the system between the flaring events to recover the free energy released by the
preceding flaring event. However, this is not the case 
as far as the flares in our observation are concerned. No indication of energy input, 
whether flux emergence or increase of magnetic shear, was detected throughout 
our observing span. We  thus speculate that  the occurrence of  a series of  flares was 
possible 
due to the complex geometry of our active region magnetic field.  A simple bipolar 
magnetic field would proceed to a lower energy state by one bursty event of 
reconnection. However,  in a  complex  active region  containing more  than  a pair  of 
magnetic poles, 
the transition to the lowest energy state may possibly comprise several steps of
macroscopic change in field topology. This speculation, of course, has to be examined 
by further studies involving many other observations and numerical experiments as well.

\section{Magnetic Forces and Linear Force-Free Field Approximation}

It is generally believed that solar magnetic fields are force-free in the corona, 
but far from it in the photosphere. However, to construct a force-free model of 
the coronal magnetic field, the field data observed at the photospheric level are 
employed as boundary conditions. Not only because a magnetohydrostatic 
equilibrium under gravity is more difficult to construct than a force-free 
solution, but also because no reliable information about plasma pressure is 
available in the photosphere, 
force-free field modeling with photospheric boundary conditions is 
being widely attempted 
(e.g., McClymont and Miki\' c, 1994 for AR 5747) 
despite the afore-mentioned inconsistency. 
The reliability of such models thus depends on how 
much the field behaves like a force-free field near the photosphere. In this 
section, we investigate the ``force-freeness'' of AR 5747.  

A force-free field is a magnetic field satisfying the Lorentz force-free condition
\begin{equation}
(\nabla \times {\bf B}) \times {\bf B} =0\, ,
\end{equation}
which can be rewritten as 
\begin{equation}
\nabla \times {\bf B} = \alpha {\bf B}\, .
\end{equation}
The so called force-free coefficient $\alpha$ is thus given by
\begin{equation}
\alpha={ J_x \over B_x}={ J_y \over B_y}={ J_z \over B_z}\, ,
\end{equation}
in rationalized electromagnetic units. 
Taking divergence of Equation (2) and using $\nabla \cdot {\bf B}=0$, we have 
\begin{equation}
{\bf B}\cdot \nabla \alpha = 0\, ,
\end{equation}
which means that $\alpha$ is a function of each field line. 
With the vector magnetogram of Oct. 20, 1989,   Canfield {\it et al.} (1991) examined 
whether the ratio of current density to field strength ($J/B$) is 
conserved along each elementary
flux tube. Although  the force-free  coefficient $\alpha=J_i/B_i$  is necessarily constant 
along each field line in a force-free field, the condition is in practice difficult to check 
due to the noises in current density in weak field regions   
(McClymont, Jiao, and Miki\'c, 1997).
To examine the force-freeness of AR 5747, 
we have computed  the integrated  Lorentz force components scaled with the integrated 
magnetic pressure force, i.e.,   $F_x/F_o$, $F_y/F_o$ and $F_z/F_o$ 
(Metcalf {\it et al.} 1995), in which 
\begin{equation}
F_x = -{1 \over 4 \pi} \int B_x B_z dx dy,
\end{equation}
\begin{equation}
F_y = -{1 \over 4 \pi} \int B_y B_z dx dy,
\end{equation}
\begin{equation}
F_z = -{1 \over 8\pi} \int (B_z^2-B_x^2-B_y^2) dx dy,
\end{equation}
and
\begin{equation}
F_o = {1 \over 8\pi} \int (B_z^2+B_x^2+B_y^2) dx dy,
\end{equation}
where $F_o$ is the integrated magnetic pressure force. 
In this calculation, only pixels with field strength larger than 100 G 
for both longitudinal  and 
transverse
fields are considered to reduce the effect of noise.
In  Table  V,   we present the normalized integrated forces obtained  from  three    
vector 
magnetograms. The absolute values of these forces are much smaller than those
at the photospheric level of Metcalf {\it et al.} (1995), which implies that our active region 
field is more or less force-free even near the solar surface.
It is also noted that the magnetic fields of AR 5747 become less force-free in a relaxation
stage as times go.

\begin{table}
\caption{Normalized integrated forces and linear force free coefficients 
for AR 5747. }
\begin{flushleft}
\begin{tabular}{cccccccc}
\hline
 & Data  &  $F_x/F_o$ &   
$F_y/F_o$ &  $F_z/F_o$ 
& $\alpha \rm [m^{-1}]$ & $\alpha^*$ &  \\
\hline
& a) &   0.008 &  0.005 & -0.126 &  -1.1$\times10^{-7}$
 &-6.8$\times10^{-8}$ & \\
 & b) &   0.042 &  0.015 & -0.148 &  -1.0$\times10^{-7}$
 &-6.1$\times10^{-8}$ & \\
 & c) &   0.097 &  0.085 & -0.222 &  -7.2$\times10^{-8}$
 &-4.5$\times10^{-8}$ & \\
\hline
\end{tabular}
\end{flushleft}
Asterisked (*) values are obtained by  minimizing the difference between  the horizontal 
components of 
the constant $\alpha$  force-free model and   the observed horizontal  magnetic fields,  
considering only pixels with $B_t > 300 \, \rm G$ (Pevtsov {\it et al.} 1996).
\end{table}

\begin{figure}
\psfig{figure=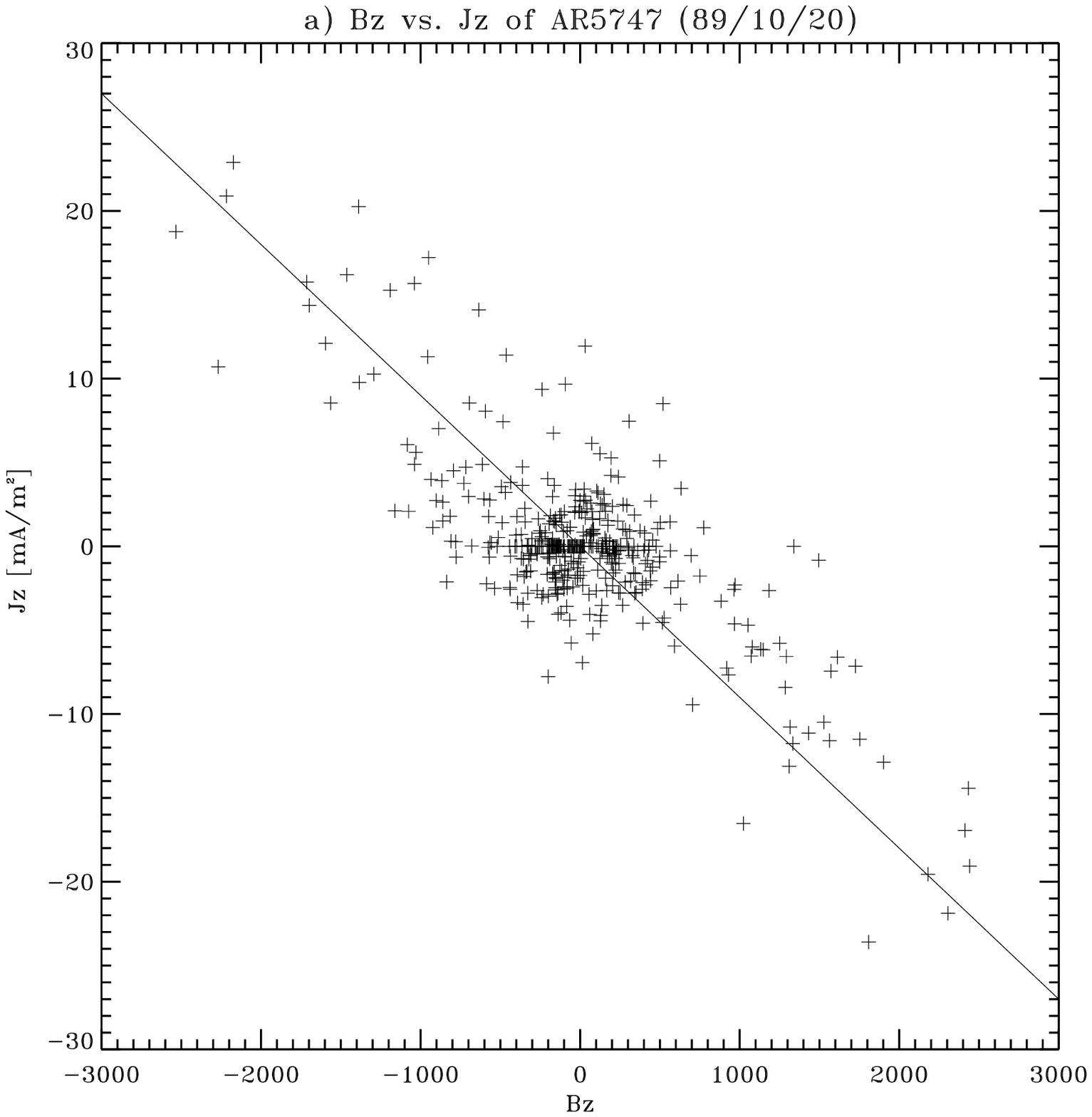,height=6cm,width=8cm}
\psfig{figure=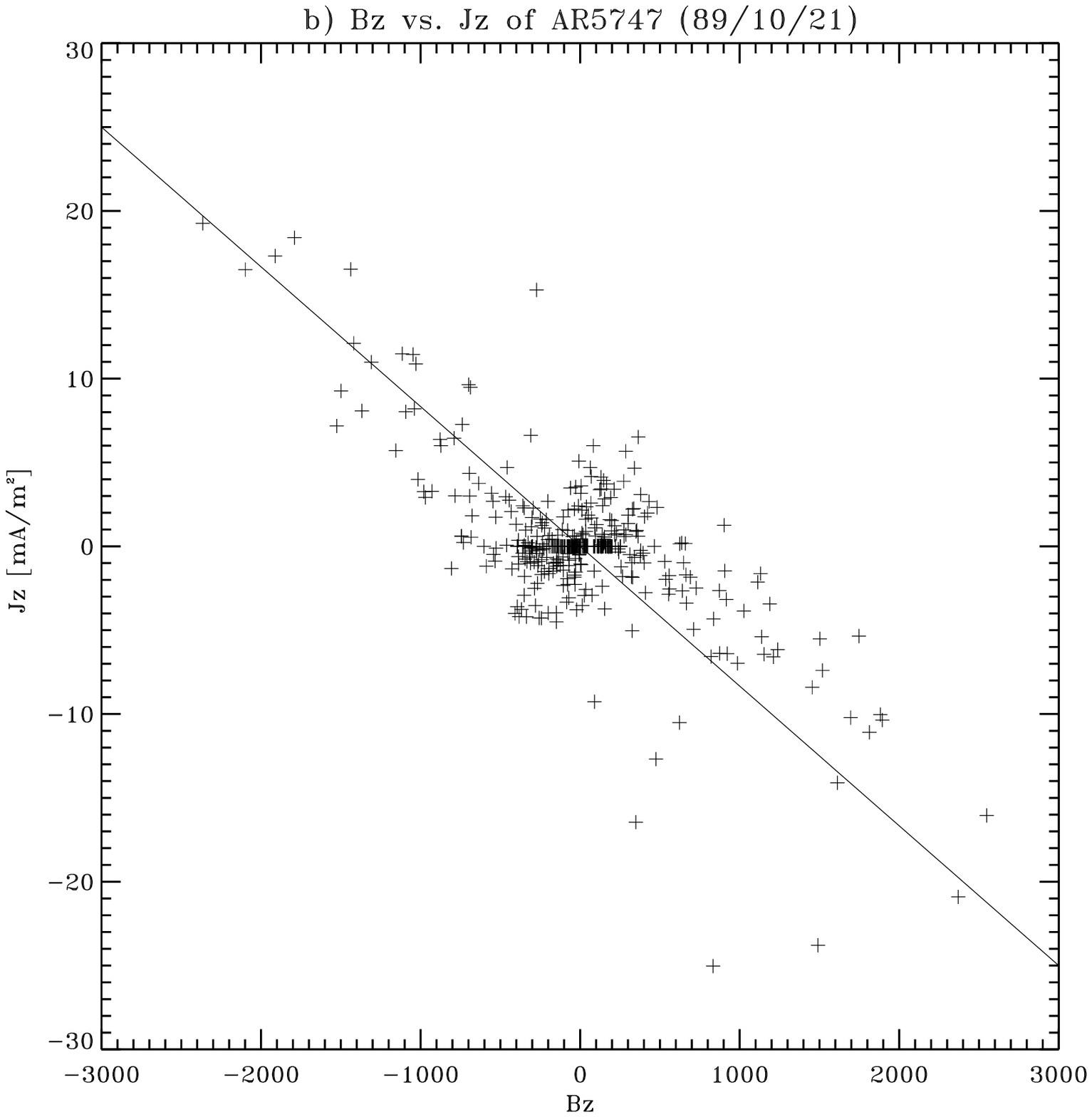,height=6cm,width=8cm}
\psfig{figure=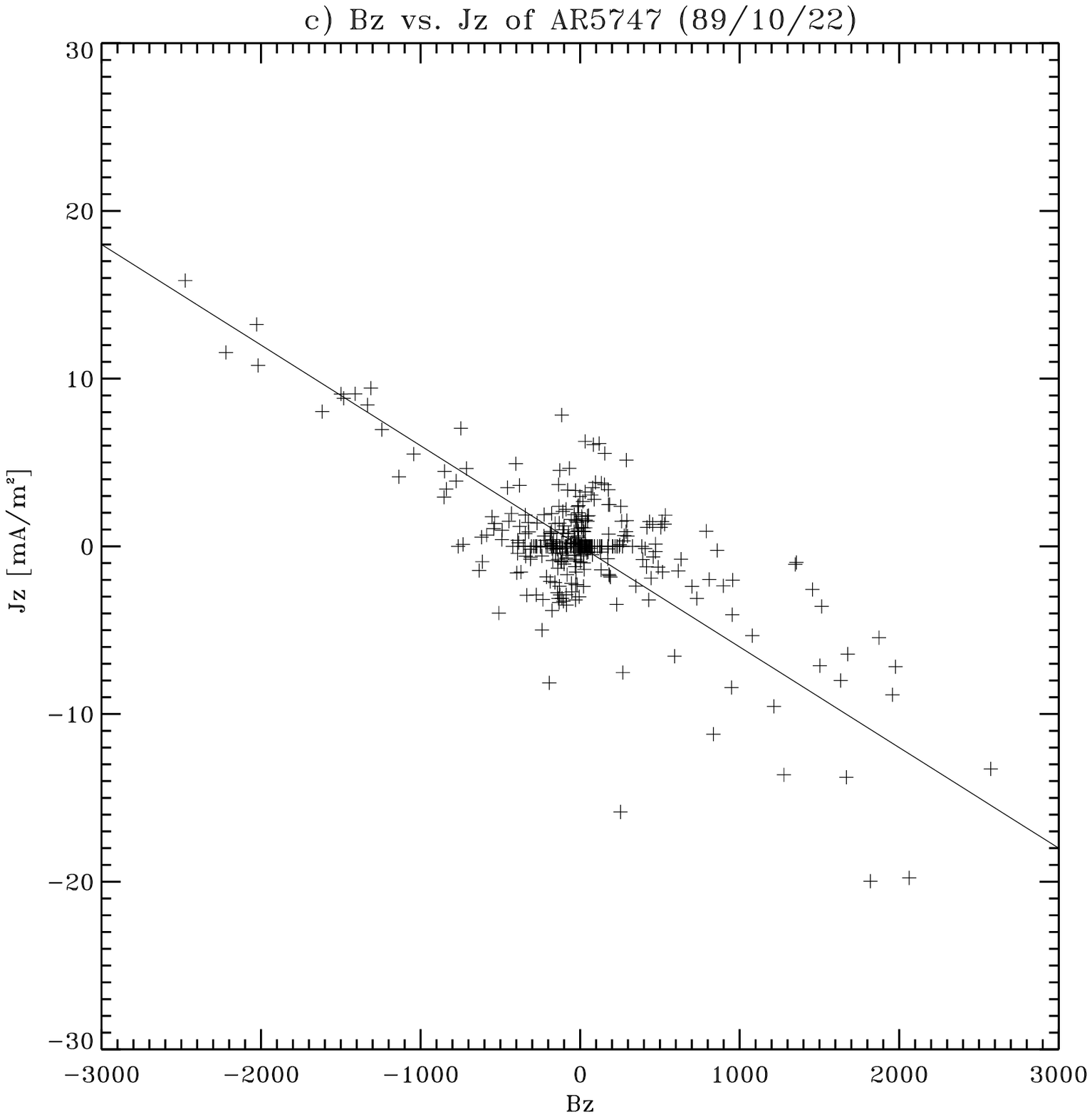,height=6cm,width=8cm}
\caption{$J_z$ vs. $B_z$ from
three vector magnetograms
of AR 5747.
Straight lines are the regression lines obtained from  eye fitting. 
}
\end{figure}

Now we turn to the question whether our active region field 
is approximately linearly 
force-free. In Figure 1, the transverse field vectors  
show a common curling pattern for each magnetic polarity, which allows us 
to expect that values of  the force-free coefficient do not  diverge much. To investigate 
the linearity, we have plotted for each data set $B_z$ vs. $J_z$ and a plausible 
regression line obtained by eye fitting in Figure 7. 
The figures show that there exist approximate linear  relationships between $B_z$ and 
$J_z$ for three vector magnetograms. 
We have already observed in Figures 1 and 2 that the distribution of vertical 
electric current density well matches that of magnetic fluxes of opposite polarity. 
In Table   V, we  have  tabulated  linear  force-free   coefficients obtained  by  linear 
regression in Figure 7. We have also listed 
the coefficients obtained by minimizing the difference between 
the horizontal components in a constant $\alpha$ force-free field model
and the horizontal field vectors in the vector magnetogram, considering 
only pixels with $B_t > 300 \, \rm G$ (Pevtsov {\it et al.}, 1996). 
In both sets, the absolute value of force-free coefficients decreased 
with time, as other nonpotentiality parameters did. 
This suggests that  the linear  force-free coefficient  could be  as good  a 
nonpotential 
evolutionary indicator  
as other nonpotentiality parameters  
as long as the linear force-free approximation is more or less valid. 
Furthermore, the linear force-free 
coefficient has a merit as a global parameter.

\section{Summary and Conclusion}

In this study,  we have  analyzed the MSO  vector magnetograms  of AR 5747 
taken on October 20 to 22,  1989.  A nonlinear least square method  was adopted 
to derive  the magnetic  field vectors  from the  observed  Stokes profiles  and a 
multi-step ambiguity solution 
method was used to resolve the
$180\arcdeg$ ambiguity.  
From the ambiguity-resolved vector magnetograms, 
we have derived a set of physical quantities which are
magnetic flux, vertical current
density, magnetic shear angle, angular shear, magnetic free energy
density and MAD, a measure of magnetic field discontinuity.  
In order to examine the force-free character of the active region field, we 
have calculated  the normalized integrated Lorentz forces and
compared the longitudinal field 
$B_z$ and the corresponding 
vertical current density $J_z$.   
Most important results from this work can  be summarized as 
follows.

1) Magnetic nonpotentiality     
is concentrated near the inversion line, where flare brightenings are observed.

2) All the physical  parameters that we have considered 
(vertical current  density, mean shear  angle, 
mean angular shear, sum of free energy density and sum of MAD)
decreased with time, which indicates that the active region was in a relaxation period.

3) The X-ray flares that occurred during the observing period could be related with
flux cancellation. Flaring events might be 
considered as bursty parts in the long term relaxation process.

4) It is found that the active 
region was approximately linearly force-free 
throughout the  observing span   
and the absolute value of the derived linear force-free 
coefficient decreased with time. 
Our result suggests that
the linear force-free coefficient 
could  be a good global parameter indicating 
the evolutionary status of an active region 
as long as the field is approximately force-free.

\acknowledgements
We wish to thank Dr. Metcalf 
 for allowing us to use some of his numerical routines
 for analyzing vector magnetograms and Dr. Pevtsov for helpful comments.
 The data from the Mees Solar Observatory, University of Hawaii 
are produced with the support of NASA grant NAG 5-4941 and
NASA contract NAS8-40801.
This work has been supported in part by
the Basic Research Fund (99-1-500-00 and 99-1-500-21) of 
Korea 
Astronomy Observatory and in part by the Korea-US Cooperative Science Program under KOSEF(995-0200-004-2).


\begin{thebibliography}{}

\bibitem[Canfield {\it et al.} 1991]{can91}
Canfield, R. C., Fan, Y. , Leka, K. D., Lites, B. W.,  and
Zirin, H.: 1991,
in 
{\it Solar Polarimetry}, Proceeding of the Workshop of Solar
Polarimetry,  ed. L. November , p. 296.

\bibitem[Canfield {\it et al.} 1993]{can93}
Canfield,   R. C.,
La Beaujardiere, J.-F., Han, Y., Leka, K. D., 
McClymont, A. N.,
Metcalf, T. R., Mickey, D. L., Wulser, J.-P., and Lites, B. W.:
1993,       
{\it Astrophys. J.} {\bf 411}, 362.


\bibitem[Hagyard {\it et al.} 1984]{hag84}
Hagyard, M. J.,  Smith, Jr., J. B., Teuber, D., and West, E. A.: 1984, {\it Solar Phys.} {\bf 91}, 115.

\bibitem[Hagyard {\it et al.} 1990]{hag90}
Hagyard, M. J., Ventkatarishnan, P., and Smith, Jr., J. B.: 1990, {\it Astrophys. J. Suppl.} {\bf 73}, 159.

\bibitem[Leka {\it et al.} 1993]{lek93}
Leka, K. D., Canfield, R. C., McClymont,  A. N., de  la Beaujardiere, J. F.,  and Fan, Y.:  
1993, 
{\it Astrophys. J.} {\bf 411}, 370.

\bibitem[McClymont and Miki\'c 1994]{mcc94}
McClymont, A. N. and Miki\'c, Z.: 1994, {\it Astrophys. J.} {\bf 422}, 899.


\bibitem[McClymont, Jiao, and Miki\'c 1997]{mcc97}
McClymont, A. N., Jiao, L., and Miki\'c, Z.: 1997, {\it Solar Phys.}
{\bf 174}, 191.

\bibitem[Metcalf {\it et al.} 1991]{met91}
Metcalf, T. R., Canfield, R. C., Mickey, D. L., and
Lites, B. W.:  1991,
in 
{\it Solar Polarimetry}, Proceeding of the Workshop of Solar
Polarimetry,  ed. L. November , p. 376 .


\bibitem[Metcalf {\it et al.} 1995]{met95}
Metcalf, T. R., Jiao, L.,  McClymont, A. N.,   Canfield,  R. C., and Uitenbroek, H.: 1995,  
{\it Astrophys. J.}
{\bf 439}, 474.


\bibitem[Mickey 1985]{mic85}
Mickey, D. L.:  1985,   
 {\it Solar Phys.}
{\bf 97}, 223.


\bibitem[Moon {\it et al.} 1999a]{moon99a}
Moon, Y.-J., Yun, H. S., Lee, S. W., Kim, J.-H. Choe, G. S., Park, Y. D., 
Ai, G., Zhang, H., and Fang, C.: 1999a, {\it Solar Phys.} {\bf 184}, 323.

\bibitem[Moon, Park, and Yun 1999b]{moon99b}
Moon, Y.-J., Park, Y. D., and Yun, H. S.:
 1999b, {\it J. Korean Astron. Soc.} {\bf 32}, 63. 

\bibitem[Moon {\it et al.} 1999c]{moon99c}
Moon, Y.-J., Yun, H. S., Choe, G. S., Park, Y. D., and Mickey, D. L.:
 1999c, submitted to Solar Physics.  (Paper I)

\bibitem[Pevtsov {\it et al.} 1996]{pev96}
Pevtsov, A. A., Canfield, R. C., and Zirin, H.:  1996,
{\it Astrophys. J.} {\bf 473}, 533.

\bibitem[Pevtsov {\it et al.} 1997]{pev97}
Pevtsov, A. A., Canfield, R. C., and McClymont, A. N.:  1997,
{\it Astrophys. J.} {\bf 481}, 973.

\bibitem[Ronan {\it et al.} 1987]{ron87}
Ronan, R. S., Mickey, D. L., and Orral, F. Q.:  1987,
{\it Solar Phys.} {\bf 113}, 353.

\bibitem[Skumanich and Lites 1987]{sku87}
Skumanich, A and Lites, B. W.: 1987, {\it Astrophys. J.} {\bf 322}, 473.


\bibitem[Wang 1997]{wan97}
Wang, H.: 1997, {\it Solar Phys.} {\bf 174}, 163.

\bibitem[Wang {\it et al.} 1996]{wan96}
Wang,  J., Shi, Z., Wang, H., and Lue, Y.: 
1996,  {\it Astrophys. J.} {\bf 456},  861.

\bibitem[Wang {\it et al.} 1994]{wan94}
Wang,  T., Xu, A., and Zhang, H.: 1994,  {\it Solar Phys.} {\bf 155}. 99.


\end{thebibliography}
\end{document}